\documentclass[11pt,a4paper]{article}
\usepackage[english]{babel}
\usepackage{amsmath,amsthm,amssymb,epsfig,latexsym}
\usepackage{color}
\usepackage[numbers,square,sort&compress]{natbib}

\newcommand{\so}{\scriptscriptstyle \rm I}
\newcommand{\st}{\scriptscriptstyle \rm I\hspace{-1pt}I}

\newcommand{\bt}{\bar t}

\newcommand{\bs}{\bar s}

\newcommand{\bu}{\bar u}
\newcommand{\bv}{\bar v}
\newcommand{\ort}{{\scriptscriptstyle\overrightarrow{\displaystyle t}}}
\newcommand{\olt}{{\scriptscriptstyle\overleftarrow{\displaystyle t}}}
\newcommand{\ors}{{\scriptscriptstyle\overrightarrow{\displaystyle s}}}
\newcommand{\ols}{{\scriptscriptstyle\overleftarrow{\displaystyle s}}}

\newcommand{\hg}{\hat\gamma}

\newcommand{\be}[1]{\begin{equation}\label{#1}}
\newcommand{\ba}[1]{\begin{multline}\label{#1}}
\newcommand{\ee}{\end{equation}}
\newcommand{\ea}{\end{eqnarray}}

\newcommand{\str}{\mathop{\rm str}}

\newtheorem{prop}{Proposition}[section]
\newtheorem{lemma}{Lemma}[section]
\newtheorem{cor}{Corollary}[section]

\def\qed{\hfill\nobreak\hbox{$\square$}\par\medbreak}
 \makeatletter
 \@addtoreset{equation}{section}
 \makeatother

\newcommand{\bea}{\begin{eqnarray}}
\newcommand{\eea}{\end{eqnarray}}

\newcommand{\bT}{\mathbb{T}}
\newcommand{\mb}[1]{\quad\mbox{#1}\quad}

\def\eps{\varepsilon}

\newcommand{\ZZ}{{\mathbb Z}}

\begin{document}

\begin{flushright}
LAPTH-010/17
\end{flushright}

\vspace{12pt}

\begin{center}
\begin{LARGE}
{\bf Scalar products of Bethe vectors\\[2mm] in the models with $\mathfrak{gl}(m|n)$ symmetry }
\end{LARGE}

\vspace{40pt}

\begin{large}
{A.~Hutsalyuk${}^{a,b}$,  A.~Liashyk${}^{c,d,e}$,
S.~Z.~Pakuliak${}^{a,f}$,\\ E.~Ragoucy${}^g$, N.~A.~Slavnov${}^h$\  \footnote{
hutsalyuk@gmail.com, a.liashyk@gmail.com, stanislav.pakuliak@jinr.ru, eric.ragoucy@lapth.cnrs.fr, nslavnov@mi.ras.ru}}
\end{large}

 \vspace{12mm}

${}^a$ {\it Moscow Institute of Physics and Technology,  Dolgoprudny, Moscow reg., Russia}

\vspace{4mm}

${}^b$ {\it Fachbereich C Physik, Bergische Universit\"at Wuppertal, 42097 Wuppertal, Germany}

\vspace{4mm}

${}^c$ {\it Bogoliubov Institute for Theoretical Physics, NAS of Ukraine,  Kiev, Ukraine}

\vspace{4mm}

${}^d$ {\it National Research University Higher School of Economics, Faculty of Mathematics, Moscow, Russia}

\vspace{4mm}

${}^e$ {\it Center for Advanced Studies, Skolkovo Institute of Science and Technology, Moscow, Russia}

\vspace{4mm}

${}^f$ {\it Laboratory of Theoretical Physics, JINR,  Dubna, Moscow reg., Russia}

\vspace{4mm}

${}^g$ {\it Laboratoire de Physique Th\'eorique LAPTh, CNRS and USMB,\\
BP 110, 74941 Annecy-le-Vieux Cedex, France}

\vspace{4mm}

${}^h$  {\it Steklov Mathematical Institute of Russian Academy of Sciences, Moscow, Russia}

\end{center}

\vspace{4mm}


\begin{abstract}
We study scalar products of Bethe vectors in the models solvable by the nested algebraic Bethe ansatz and described by  $\mathfrak{gl}(m|n)$
superalgebra. Using coproduct  properties of the Bethe vectors we  obtain a {\it sum formula} for their scalar products.
This formula describes the scalar product in terms  of a sum over partitions of Bethe parameters. We also obtain recursions for the
Bethe vectors. This allows us to find recursions for the highest coefficient of the scalar product.
\end{abstract}

\vspace{1cm}

\section{Introduction}

The problem of calculating correlation functions of quantum exactly solvable models is of great importance. The creation of the Quantum Inverse Scattering
Method (QISM) in the early 80s of the last century provided a powerful tool for investigating this problem \cite{FadST79,FadT79,BogIK93L,FadLH96}.
The first works in which QISM was applied to the problem of correlation functions \cite{Kor82,IzeK84} were devoted to the models related to the different deformations of the affine algebra $\widehat{\mathfrak{gl}}(2)$. Already in those papers,  the key role  of Bethe vectors  scalar products
was established. In particular, a {\it sum formula} for the scalar product of  Bethe vectors was obtained in \cite{Kor82}. This formula gives the scalar product as a sum over partitions of Bethe parameters.

A generalization of QISM to the models with higher rank symmetry was given in papers \cite{KulRes81,KulRes83,KulRes82} where the nested algebraic Bethe ansatz was developed. There a recursive procedure was developed  to  construct Bethe vectors corresponding to the  $\mathfrak{gl}(N)$ algebra from the known Bethe vectors of the $\mathfrak{gl}(N-1)$ algebra. The problem of the scalar products in $SU(3)$-invariant models were studied in \cite{Res86}, where  an analog of the sum formula for the scalar product was obtained and the norm of the transfer matrix eigenstates was computed. Recently in a series of
papers \cite{Whe12,Whe13,BelPRS12a,BelPRS12b,HutLPRS16c,HutLPRS16b} the Bethe vectors scalar products in the models with ${\mathfrak{gl}}(3)$ and
${\mathfrak{gl}}(2|1)$ symmetries were intensively studied. There determinant representations for some important particular cases were obtained leading
eventually to the determinant formulas for form factors of local operators in the corresponding physical models \cite{PakRS15a,PakRS15c,HutLPRS16d,FukS17}.
A generalization of some of those results to the models with trigonometric $R$-matrix was given in \cite{PakRS14a,Sla15a}.

Concerning the scalar products in the models with higher rank (super) symmetries, only few results are known for today. First, it is worth
mentioning the papers \cite{VT,TarV96}, in which the authors developed a new approach to the problem based on the quantized Knizh\-nik--Za\-mo\-lod\-chi\-kov equation. There the norms of the transfer matrix eigenstates in $\mathfrak{gl}(N)$-based models were calculated.
 Some partial results were also obtained when specializing to fundamental representations or to particular cases of Bethe vectors
\cite{MukV05,WheelF13,EscGSV11,Grom16}.

In this paper we study the Bethe vectors scalar products in the models described by $\mathfrak{gl}(m|n)$ superalgebras. Hence it encompasses the case of $\mathfrak{gl}(m)$ algebras.
In spite of we work within the framework of the traditional approach based on the nested algebraic Bethe ansatz, we essentially use recent results obtained in \cite{HutLPRS17a} via the method
of projections for construction of Bethe vectors.  This method was proposed in the paper \cite{KhP-Kyoto}. It uses the relation between two different realizations of the quantized Hopf algebra $U_q(\widehat{\mathfrak{gl}}(N))$ associated with the
affine algebra $\widehat{\mathfrak{gl}}(N)$, one in terms of the universal monodromy matrix $T(z)$ and $RTT$-commutation relations and second in terms of the
total currents, which are defined by the Gauss decomposition of the monodromy matrix $T(z)$ \cite{DiFr93}. In \cite{HutLPRS17a} we generalized this approach
to the case of the Yangians of $\mathfrak{gl}(m|n)$ superalgebras. Among the results of \cite{HutLPRS17a} that are used in the present paper, we note the formulas for the action of the monodromy matrix entries onto the Bethe vectors, and also the coproduct formula for the Bethe vectors.

The main result of this paper is the sum formula for the scalar product of Bethe vectors. In our previous publications  (see e.g. \cite{HutLPRS16c,PakRS14a}) we derived it using explicit formulas  of the
monodromy matrix elements   multiple actions onto the Bethe vectors. This method is straightforward, but it becomes rather cumbersome already for $\mathfrak{gl}(3)$ and $\mathfrak{gl}(2|1)$ based models. Furthermore, the possibility of its application to the models with higher rank symmetries is
under question. Instead, in the present paper we use a method based on the coproduct formula for the Bethe vectors. Actually, the structure of the
scalar product is encoded in the coproduct formula. Therefore, this method directly leads to the sum formula, in which the scalar product is given as
a sum over partitions of Bethe parameters.

The sum formula contains an important object called the highest coefficient (HC) \cite{Kor82}. In the $\mathfrak{gl}(2)$ based models and their
$q$-deformation the HC coincides with a partition function of the six-vertex model with domain wall boundary condition. An explicit representation
for it was found in \cite{Ize87}. In the models with $\mathfrak{gl}(3)$ symmetry the HC also can be associated with a special partition function, however, its
explicit form is much more sophisticated (see e.g. \cite{Whe12,BelPRS12a}). One can expect that in the case of higher rank algebras an analogous
explicit formula for the HC becomes too complex. Therefore, in this paper we do not derive such formulas, but instead, we obtain recursions,
which allow one to construct  the HC starting with the ones in the models with lower rank symmetries. These recursions can be derived from recursions
on the Bethe vectors that we also obtain in this paper.

As we have already mentioned, the Bethe vectors scalar products  are of great importance in the problem of correlation functions of quantum integrable
models. Certainly, the sum  formula is not convenient for its direct applications, as it contains a big number of terms, which grows exponentially in the
thermodynamic limit. However, it gives a key for studying particular cases of scalar products, in which the sum over partitions
can be reduced to a single determinant. This type of formulas can be used for calculating form factors of various integrable models of physical interest,
like, for instance, the Hubbard model \cite{EssK94}, the t-J model \cite{For89,EssK92,FoeK93} or multi-component Bose/Fermi gas \cite{Sch87}, not to mention spin chain models as they are nowadays tested in condensed matter experiments \cite{BachFoer}.
We also hope that our results will be of some interest in the context of super-Yang-Mills theories, when studied in the integrable systems framework.
Indeed, in these theories, the general approach relies on a spin chain based on the $\mathfrak{psu}(2,2|4)$ superalgebra.
We believe that the present results will contribute to a better understanding of the theory.

The article is organized as follows. In section~\ref{S-N} we introduce the model under consideration. There we
also specify our conventions and notation. In section~\ref{S-BV} we describe Bethe vectors of $\mathfrak{gl}(m|n)$-based
models.
Section~\ref{S-MR} contains the main results of the paper. Here
we give a sum formula for the scalar product of generic Bethe vectors and recursion relations for the Bethe vectors and the highest
coefficient. The rest of the paper contains the proofs of the results announced in section~\ref{S-MR}.
In section~\ref{S-RfBV} we prove recursion formulas for the Bethe vectors.
Section~\ref{S-SF} contains a proof of the sum formula for the scalar product.
In section~\ref{S-HC} we  study highest coefficient and find a recursion for it.
Proofs of some auxiliary statements are gathered in appendices.

\section{Description of the model\label{S-N}}

\subsection{$\mathfrak{gl}(m|n)$-based models\label{SS-glmnbm}}

The $R$-matrix of $\mathfrak{gl}(m|n)$-based models acts in the tensor product $\mathbf{C}^{m|n}\otimes \mathbf{C}^{m|n}$,
where $\mathbf{C}^{m|n}$ is the $\mathbb{Z}_2$-graded vector space with the grading $[i]=0$ for $1\le i\le m$, $[i]=1$ for $m<i\le m+n$. Here, we assume that $m\ge 1$ and $n\ge 1$, but we want to stress that our considerations are applicable to the case $m=0$ or $n=0$ as well,
i.e. to the non-graded algebras.
 Matrices acting in this space are also graded. We define this grading on the basis  of elementary units $E_{ij}$
 as $[E_{ij}]=[i]+[j]\in\ZZ_2$  (recall that $(E_{ij})_{ab}=\delta_{ia}\delta_{jb}$). The tensor products
of $\mathbf{C}^{m|n}$ spaces are  graded as follows:
\be{tens-prod}
(\mathbf{1}\otimes E_{ij})\,\cdot\,(E_{kl}\otimes \mathbf{1}) = (-1)^{([i]+[j])([k]+[l])}\,E_{kl}\otimes E_{ij}.
\ee

The $R$-matrix of $\mathfrak{gl}(m|n)$-invariant models has the form
 \be{R-mat}
 R(u,v)=\mathbb{I}+g(u,v)P, \qquad g(u,v)=\frac{c}{u-v}.
 \ee
Here $c$ is a constant,  $\mathbb{I}$ and $P$ respectively are the identity matrix and the graded permutation operator \cite{KulS80}:
\be{EP}
\mathbb{I}=\mathbf{1}\otimes\mathbf{1}=\sum_ {i,j=1}^{n+m}E_{ii}\otimes E_{jj}, \qquad P=\sum_ {i,j=1}^{n+m}(-1)^{[j]}E_{ij}\otimes E_{ji}.
\ee

The key object of QISM is a quantum monodromy matrix $T(u)$. Its matrix  elements $T_{i,j}(u)$  are graded in the same way as the matrices $[E_{ij}]$: $[T_{i,j}(u)]=[i]+[j]$. The grading is a morphism, i.e.
$[T_{i,j}(u)\cdot T_{k,l}(v)]=[T_{i,j}(u)]+[T_{k,l}(v)]$. Their commutation relations
are given by the $RTT$-relation
\be{RTT}
R(u,v)\bigl(T(u)\otimes \mathbf{1}\bigr) \bigl(\mathbf{1}\otimes T(v)\bigr)= \bigl(\mathbf{1}\otimes T(v)\bigr)
\bigl( T(u)\otimes \mathbf{1}\bigr)R(u,v).
\ee
Equation \eqref{RTT} holds in the tensor product $\mathbf{C}^{m|n}\otimes \mathbf{C}^{m|n}\otimes\mathcal{H}$,
where $\mathcal{H}$ is a Hilbert space of the Hamiltonian  under consideration. Here all the tensor products are graded.

The $RTT$-relation \eqref{RTT} yields a set of  commutation relations for the monodromy matrix elements
\begin{equation}\label{TM-1}
\begin{split}
[T_{i,j}(u),T_{k,l}(v)\}&
=(-1)^{[i]([k]+[l])+[k][l]}g(u,v)\Big(T_{k,j}(v)T_{i,l}(u)-T_{k,j}(u)T_{i,l}(v)\Big)\\
&=(-1)^{[l]([i]+[j])+[i][j]}g(u,v)\Big(T_{i,l}(u)T_{k,j}(v)-T_{i,l}(v)T_{k,j}(u)\Big),
\end{split}
\end{equation}
where we introduced the graded commutator
\be{Def-SupC}
[T_{i,j}(u),T_{k,l}(v)\}= T_{i,j}(u)T_{k,l}(v) -(-1)^{([i]+[j])([k]+[l])}   T_{k,l}(v)  T_{i,j}(u).
\ee

The graded transfer matrix is defined as the supertrace of the monodromy matrix
\be{transfer.mat}
\mathcal{T}(u)=\str T(u)= \sum_{j=1}^{m+n} (-1)^{[j]}\, T_{j,j}(u).
\ee
One can easily check \cite{KulS80} that $[\mathcal{T}(u)\,,\,\mathcal{T}(v)]=0$. Thus the transfer matrix can be used as
a generating function of integrals of motion of an integrable system.

\subsection{Notation}

In this paper we use notation and conventions of the work \cite{HutLPRS17a}.
Besides the function $g(u,v)$  we introduce  two rational functions
\be{fh}
\begin{aligned}
f(u,v)&=1+g(u,v)=\frac{u-v+c}{u-v},\\
h(u,v)&=\frac{f(u,v)}{g(u,v)}=\frac{u-v+c}{c}.
\end{aligned}
\ee
In order to make formulas uniform we also use `graded' functions
\be{desand}
\begin{aligned}
g_{[i]}(u,v)&=(-1)^{[i]}g(u,v)=\frac{(-1)^{[i]}c}{u-v},\\
f_{[i]}(u,v)&=1+g_{[i]}(u,v)=\frac{u-v+(-1)^{[i]}c}{u-v},\\
h_{[i]}(u,v)&=\frac{f_{[i]}(u,v)}{g_{[i]}(u,v)}=\frac{(u-v)+(-1)^{[i]}c}{(-1)^{[i]}c},
\end{aligned}
\ee
and
\be{gam-hg}
\gamma_{i}(u,v)=\frac{f_{[i]}(u,v)}{h(u,v)^{\delta_{i,m}}},\qquad \hg_{i}(u,v)=\frac{f_{[i+1]}(u,v)}{h(v,u)^{\delta_{i,m}}}.
\ee
%
%
%
Observe that we use the subscript $i$ for the functions $\gamma$ and $\hg$ instead of the subscript $[i]$. This is because
these functions actually take three values. For example, $\gamma_i(u,v)=f(u,v)$ for $i<m$, $\gamma_i(u,v)=g(u,v)$ for $i=m$,
and $\gamma_i(u,v)=f(v,u)$ for $i>m$.
It is also easy to see that $\hg_{i}(u,v)=(-1)^{\delta_{i,m}}\gamma_{i}(u,v)$.

Let us formulate now a convention on the notation.
We  denote sets of variables by bar, for example, $\bu$. When dealing with several of them, we may
equip these sets or subsets with additional superscript: $\bs^i$, $\bt^\nu$, etc.
Individual elements of the sets or subsets are denoted by Latin subscripts, for instance,
$u_j$ is an element of $\bu$, $t^i_k$ is an element of $\bt^i$ etc. As a rule, the number of elements in the
sets is not shown explicitly in the equations, however we give these cardinalities in
special comments to the formulas.
We assume that the elements in every subset of variables are ordered in such a way that the sequence of
their subscripts is strictly increasing: $\bt^i=\{t^i_1,t^i_2,\dots,t^i_{r_i}\}$. We call this ordering the natural order.

We use a shorthand notation for products of  the rational functions \eqref{fh}--\eqref{gam-hg}.
 Namely, if some of these functions depends
on a set of variables (or two sets of variables), this means that one should take the product over the corresponding set (or
double product over two sets).
For example,
 \be{SH-prod}
 \begin{aligned}
 g(\bu,v)&= \prod_{u_j\in\bu} g(u_j,v),\\
 f_{[i]}(t^{i-1}_k,\bt^i)&=\prod_{t^i_\ell\in\bt^i} f_{[i]}(t^{i-1}_k,t^i_\ell),\\
  \gamma_{\ell}(\bs^i,\bt^\ell)&=\prod_{s^i_j\in\bs^i}\prod_{t^\ell_k\in\bt^\ell} \gamma_{\ell}(s^i_j,t^\ell_k).
  \end{aligned}
 \ee
By definition, any product over the empty set is equal to $1$. A double product is equal to $1$ if at least one of the sets is empty.

Below we will extend this convention to the  products of monodromy matrix entries and their
 eigenvalues (see \eqref{SH-prod1} and \eqref{SH-prod2}).

\section{Bethe vectors \label{S-BV}}

Bethe vectors belong to the space $\mathcal{H}$ in which the monodromy matrix entries act. We do not specify this space,
however, we assume that it contains a {\it pseudovacuum vector} $|0\rangle$, such that
 \be{Tij}
 \begin{aligned}
 &T_{i,i}(u)|0\rangle=\lambda_i(u)|0\rangle, &\qquad i&=1,\dots,m+n,\\
 & T_{i,j}(u)|0\rangle=0, &\qquad i&>j\,,
 \end{aligned}
 \ee
where $\lambda_i(u)$ are some scalar functions. In the framework of the generalized model \cite{Kor82} considered in this paper,
they remain free functional parameters. Below it will be convenient to deal with ratios of these functions
 \be{ratios}
 \alpha_i(u)=\frac{\lambda_i(u)}{\lambda_{i+1}(u)}, \qquad  i=1,\dots,m+n-1.
 \ee

We extend the convention on the shorthand notation \eqref{SH-prod} to the products of the functions introduced above, for example,
 \be{SH-prod1}
 \lambda_k(\bu)= \prod_{u_j\in\bu} \lambda_k(u_j),\qquad
 \alpha_i(\bt^i)=\prod_{t^i_\ell\in\bt^i}  \alpha_i(t^i_\ell).
 \ee
We use the same convention for the products of commuting operators
 \be{SH-prod2}
T_{i,j}(\bu)= \prod_{u_j\in\bu} T_{i,j}(u_j),\qquad\text{for}\qquad [i]+[j]=0,\quad \mod 2.
\ee
Finally, for the product of odd operators $T_{i,j}$ with $[i]+[j]=1$ we introduce a special notation
\be{bbT}
\begin{aligned}
&\bT_{i,j}(\bu)=\frac{T_{i,j}(u_1)\dots T_{i,j}(u_p)}{\prod_{1\le k<\ell\le p}h(u_\ell,u_k)},\qquad [i]+[j]=1,\qquad i<j,\\
&\bT_{i,j}(\bu)=\frac{T_{i,j}(u_1)\dots T_{i,j}(u_p)}{\prod_{1\le k<\ell\le p}h(u_k,u_\ell)},\qquad [i]+[j]=1,\qquad i>j.
\end{aligned}
\ee
Due to the commutation relations \eqref{TM-1} the operator products \eqref{bbT} are symmetric over permutations of the
parameters $\bu$.

\subsection{Coloring\label{SS-C}}
In physical models, vectors of the space $\mathcal{H}$  describe states with quasiparticles of different types (colors). In
$\mathfrak{gl}(m|n)$-based models quasiparticles may have $N=m+n-1$ colors. Let $\{r_1,\dots,r_N\}$ be a set of non-negative integers.
We say that a state has coloring $\{r_1,\dots,r_N\}$, if it contains $r_i$ quasiparticles of the color $i$.
A state with a fixed coloring can be obtained by successive application of the creation operators $T_{i,j}$ with $i<j$
to the vector $|0\rangle$, which has zero coloring. Acting on this state, an operator $T_{i,j}$ adds quasiparticles with the colors $i,\dots, j-1$, one particle of each color. In particular, the operator $T_{i,i+1}$ creates one quasiparticle of the color $i$, the operator $T_{1,n+m}$ creates $N$ quasiparticles of $N$ different colors. The diagonal operators $T_{i,i}$ are neutral, the matrix elements $T_{i,j}$ with $i>j$ play the role of annihilation operators.
Acting on the state of a fixed  coloring, the annihilation operator $T_{i,j}$ removes from this state the quasiparticles with the colors $j,\dots, i-1$, one particle of each color. In particular,  if $j-1 <k <i$, and the annihilation operator $T_{i,j}$ acts on a state in which there is no
particles of the color $k$, then this action vanishes.

This definition can be formalized at the level of the Yangian through the Cartan generators of the Lie superalgebra  $\mathfrak{gl}(m|n)$. Indeed, the zero modes
$$
T_{ij}[0]=\lim_{u\to\infty} \frac{u}c\big(T_{ij}(u)-\delta_{ij}\big)$$
form a $\mathfrak{gl}(m|n)$ superalgebra, with commutation relations
\be{}
[T_{ij}[0]\,,\, T_{kl}[0]\} = (-1)^{[i]([k]+[l])+[k][l]}\Big( \delta_{il}\,T_{kj}[0]-\delta_{jk}\,T_{il}[0] \Big)\,,\quad i,j,k,l=1,...,m+n.
\ee
This superalgebra is a symmetry of the generalized model, since it commutes with the transfer matrix, $[T_{ij}[0]\,,\,{\cal T}(z)]=0$, $i,j=1,...,m+n$. In fact the monodromy matrix entries form a representation of this superalgebra:
\be{}
[T_{ij}[0]\,,\, T_{kl}(z)\} = (-1)^{[i]([k]+[l])+[k][l]}\Big(\delta_{il}\,T_{kj}(z)-\delta_{jk}\,T_{il}(z) \Big)\,,\quad i,j,k,l=1,...,m+n.
\ee
In particular, for the Cartan generators $T_{jj}[0]$ we obtain
\be{}
{[T_{jj}[0]\,,\, T_{kl}(z)]} = (-1)^{[j]}\big(\delta_{jl}-\delta_{jk} \big)\,T_{kl}(z)\,,\quad j,k,l=1,...,m+n.
\ee

Then, the colors correspond to the eigenvalues under the Cartan generators\footnote{The last generator $h_{m+n}$ is central, see \eqref{color}.}
\be{}
h_j=\sum_{k=1}^j (-1)^{[k]}\,T_{kk}[0]\,,\quad j=1,...,m+n-1.
\ee
Indeed, one can check that
\be{color}
[h_j\,,\,T_{kl}(z)] = \eps_j(k,l)\, T_{kl}(z) \mb{with}
\begin{cases} \eps_j(k,l)=-1 \mb{if} k\leq j<l \\ \eps_j(k,l)=+1 \mb{if} l\leq j<k \\ \eps_j(k,l)=0 \mb{otherwise}  \end{cases}
\ee
These eigenvalues just correspond to creation/annihilation operators as described above.

Bethe vectors are certain polynomials in the creation operators $T_{i,j}$ applied to the vector $|0\rangle$.
Since Bethe vectors are eigenvectors under the Cartan generators $T_{kk}[0]$, they are also eigenvectors of the color generators $h_j$,
and hence contain only terms with the same coloring.

\textsl{Remark.} In various models of physical interest the coloring of the Bethe vectors obeys certain constraints, for instance,
$r_1\ge r_2\ge\dots\ge r_N$. In particular, this case occurs if the monodromy matrix of the model is given by the product of the
$R$-matrices \eqref{R-mat} in the fundamental representation. We do not restrict ourselves with this particular case and do not impose
any restriction for the coloring of the Bethe vectors. Thus, in what follows $r_i$ are arbitrary non-negative integers.

In this paper we do not use an explicit form of the Bethe vectors, however, the reader can find it
in \cite{HutLPRS17a}. A generic Bethe vector of $\mathfrak{gl}(m|n)$-based model depends on $N=m+n-1$ sets
of variables $\bt^1,\bt^2,\dots,\bt^N$ called Bethe parameters. We denote Bethe vectors by $\mathbb{B}(\bt)$, where
\begin{equation}\label{Bpar}
\bar{t}\
=\{t^{1}_{1},\dots,t^{1}_{r_{1}};
t^{2}_{1},\dots,t^{2}_{r_{2}};\dots;
t^{N}_{1},\dots,t^{N}_{r_{N}} \},
\end{equation}
and the  cardinalities $r_i$  of the sets $\bt^i$ coincide with the coloring. Thus, each Bethe parameter $t^i_k$ can
be associated with a quasiparticle of the color $i$.

Bethe vectors are symmetric
over permutations of the parameters $t^i_k$ within the set $\bt^i$, however, they are not symmetric over
permutations over parameters belonging to different sets $\bt^i$ and $\bt^j$. For generic Bethe vectors the
Bethe parameters $t^i_k$ are generic complex numbers. If these parameters satisfy a special system of equations
(Bethe equations), then the corresponding vector becomes an eigenvector of the transfer matrix \eqref{transfer.mat}.
In this case it is called {\it on-shell Bethe vector}. In this paper we  consider generic Bethe vectors,
however, some formulas (for instance, the sum formula for the scalar product \eqref{al-dep}, \eqref{HypResh}) can be specified to the case of
on-shell Bethe vectors as well.

Though we do not use the explicit form of the Bethe vectors, we should fix their normalization. We have already mentioned  that
a generic Bethe vector has the form of a polynomial in $T_{i,j}$ with $i<j$ applied to the pseudovacuum $|0\rangle$. Among all the terms
of this polynomial there is one monomial that contains the operators $T_{i,j}$ with $j-i=1$ only. Let us call this term the {\it main term} and denote
it by $\widetilde{\mathbb{B}}(\bt)$. Then
\be{mainterm-def}
\mathbb{B}(\bt)=\widetilde{\mathbb{B}}(\bt)+\dots.
\ee
where ellipsis means all the terms containing at least one operator $T_{i,j}$ with $j-i>1$.
We will fix the normalization of the Bethe vectors by fixing a numeric coefficient
of the main term
\be{Norm-MT}
\widetilde{\mathbb{B}}(\bt)=
\frac{ \bT_{1,2}(\bt^1)\dots \bT_{N,N+1}(\bt^N)|0\rangle}
{\prod_{i=1}^{N}\lambda_{i+1}(\bt^{i})\prod_{i=1}^{N-1}  f_{[i+1]}(\bt^{i+1},\bt^i)},
\ee
where
\be{bT}
\bT_{i,i+1}(\bt^i)=\frac{T_{i,i+1}(t^i_1)\dots T_{i,i+1}(t^i_{r_i})}{\Bigl(\prod_{1\le j<k\le r_i}h(t^i_k,t^i_j)\Bigr)^{\delta_{i,m}}}.
\ee
Recall that we use here the shorthand notation for
the products of the functions $\lambda_{j+1}$ and $f_{[j+1]}$.
The  normalization in \eqref{Norm-MT} is different from the one used in \cite{HutLPRS17a} by the product $\prod_{j=1}^{N}\lambda_{j+1}(\bt^{j})$.
This additional normalization factor is convenient, because in this case the scalar products of the Bethe vectors depend
on the ratios $\alpha_i$ \eqref{ratios} only.

Since the operators $T_{i,i+1}$ and $T_{j,j+1}$ do not commute for $i\ne j$, the main term can be written in several forms corresponding to different
ordering of the monodromy matrix entries. The ordering  in \eqref{Norm-MT} naturally arises if we construct Bethe vectors via the embedding  of
$\mathfrak{gl}(m-1|n)$ to $\mathfrak{gl}(m|n)$.

\subsection{Morphism of Bethe vectors\label{SS-MBV}}

Yangians $Y(\mathfrak{gl}(m|n))$ and $Y(\mathfrak{gl}(n|m))$ are related by a morphism $\varphi$ \cite{PakRS17}
\begin{equation}\label{morphism}
\varphi:\ \begin{cases}
Y(\mathfrak{gl}(m|n))&\to\qquad Y(\mathfrak{gl}(n|m)),\\
\quad T^{m|n}_{i,j}(u) &\to\ (-1)^{[i][j]+[j]+1}\,T^{n|m}_{N+2-j,N+2-i}(u)\,,\qquad i,j=1,\dots,N+1,
\end{cases}
\end{equation}
and we recall that $N=m+n-1$. Here we also have equipped the operators $T_{ij}$ with additional superscripts showing the corresponding Yangians. This mapping
also acts on the vacuum eigenvalues $\lambda_i(u)$ \eqref{Tij} and their ratios $\alpha_i(u)$ \eqref{ratios}
\begin{equation}\label{morphism-1}
\varphi:\ \begin{cases}
\quad \lambda_{i}(u) &\to\ -\lambda_{N+2-i}(u),\qquad i=1,\dots,N+1\,,\\
\quad \alpha_{i}(u) &\to\ \frac{1}{\alpha_{N+1-i}(u)}\,,\qquad i=1,\dots,N\,.
\end{cases}
\end{equation}

Morphism $\varphi$ induces a mapping of Bethe vectors $\mathbb{B}^{m|n}$ of $Y(\mathfrak{gl}(m|n))$ to  Bethe
vectors $\mathbb{B}^{n|m}$ of $Y(\mathfrak{gl}(n|m))$. To describe this mapping
we introduce special orderings of the sets of Bethe parameters. Namely, let
\be{orders}
\ort=\{\bt^1,\bt^{2},\dots, \bt^{N}\} \qquad\text{and}\qquad \olt= \{\bt^{N},\dots,\bt^{2},\bt^{1}\}.
\ee
The ordering of the Bethe parameters within every set $\bt^k$ is not essential. Then
\be{morph-BV}
\varphi\Bigl(\mathbb{B}^{m|n}(\ort)\Bigr)=
\frac{(-1)^{r_m}\mathbb{B}^{n|m}(\olt)}
{\prod_{k=1}^N\alpha_{N+1-k}(\bt^k)}.
\ee

Applying the mapping \eqref{morph-BV} to $\mathbb{B}^{m|n}$ and then replacing $m\leftrightarrow n$ we obtain an alternative description
of the Bethe vectors corresponding to the embedding of $\mathfrak{gl}(m|n-1)$ to $\mathfrak{gl}(m|n)$.
The use of $\varphi$ \eqref{morph-BV} allows one to establish important properties of the Bethe vectors scalar products
(see section~\ref{SS-SHC}).

\subsection{Dual Bethe vectors\label{SS-MoBV}}

Dual Bethe vectors belong to the dual space $\mathcal{H}^*$, and they are
polynomials in $T_{i,j}$ with $i>j$ applied from the right to the dual pseudovacuum vector $\langle0|$.
This vector possesses properties similar to \eqref{Tij}
 \be{dTij}
 \begin{aligned}
 &\langle0|T_{i,i}(u)=\lambda_i(u)\langle0|,&\qquad i&=1,\dots,m+n,\\
 & \langle0|T_{i,j}(u)=0\,,&\qquad\qquad i&<j\,,
 \end{aligned}
 \ee
where the functions $\lambda_i(u)$ are the same as in \eqref{Tij}.

We denote dual Bethe vectors by $\mathbb{C}(\bt)$, where the set of Bethe parameters $\bt$ consists of
several sets $\bt^i$ as in \eqref{Bpar}. Similarly to how it was done for Bethe vectors, we can introduce the coloring of the dual Bethe vectors.
At the same time the role of creation and annihilation operators are reversed.

One can obtain dual Bethe vectors via a special antimorphism of the algebra \eqref{RTT} \cite{PakRS17}
\begin{equation}\label{antimo}
\Psi:\ T_{i,j}(u) \to (-1)^{[i]([j]+1)}T_{j,i}(u).
\end{equation}
This antimorphism is nothing but a super  (or equivalently, graded) transposition compatible with
the notion of supertrace. It satisfies a property
\begin{equation}\label{pro-anti}
\Psi(A\cdot B)=(-1)^{[A][B]}\Psi(B)\cdot\Psi(A),
\end{equation}
where  $A$ and $B$ are arbitrary elements of the monodromy matrix. If we extend the action of this
antimorphism to the pseudovacuum vectors by
\begin{equation}\label{Avac}
\begin{aligned}
&\Psi\bigl(|0\rangle\bigr)=\langle0|,\qquad & \Psi\bigl(A|0\rangle\bigr)=\langle0|\Psi\bigl(A\bigr),\\
&\Psi\bigl(\langle0|\bigr)=|0\rangle,\qquad& \Psi\bigl(\langle0|A\bigr)=\Psi\bigl(A\bigr)|0\rangle,
\end{aligned}
\end{equation}
then it turns out that \cite{HutLPRS17a}
\be{antimoBV}
\Psi\bigl(\mathbb{B}(\bt)\bigr)=\mathbb{C}(\bt), \qquad
\Psi\bigl(\mathbb{C}(\bt)\bigr)=(-1)^{r_m}\mathbb{B}(\bt),
\ee
where $r_m=\#\bt^m$.

{\sl Remark.} It should not be surprising that $\Psi^2\bigl(\mathbb{B}(\bt)\bigr)\ne\mathbb{B}(\bt)$. The point is that the antimorphism
$\Psi$ is idempotent of order $4$ and its square is the parity operator (counting the number of odd
monodromy matrix elements modulo $2$).

Thus, dual Bethe vectors are polynomials in $T_{i,j}$ with $i>j$ acting from the right onto $\langle0|$.
They also contain the main term $\widetilde{\mathbb{C}}(\bt)$,
which now consists of the operators $T_{i,j}$ with $i-j=1$. The main term of the dual Bethe vector can be obtained from \eqref{Norm-MT}
via the mapping $\Psi$:
\be{dNOrm1}
\widetilde{\mathbb{C}}(\bt)=\frac{(-1)^{r_m(r_m-1)/2}
\langle0|\bT_{N+1,N}(\bt^N)\dots \bT_{2,1}(\bt^1)}
{ \prod_{i=1}^{N}\lambda_{i+1}(\bt^{i})\prod_{i=1}^{N-1}  f_{[i+1]}(\bt^{i+1},\bt^i)},
\ee
where
\be{dbT}
\bT_{i+1,i}(\bt^i)=\frac{T_{i+1,i}(t^i_1)\dots T_{i+1,i}(t^i_{r_i})}{\Bigl(\prod_{1\le j<k\le r_i}h(t^i_j,t^i_k)\Bigr)^{\delta_{i,m}}}.
\ee

Finally, using the morphism $\varphi$ we obtain a relation between dual Bethe vectors corresponding to the Yangians $Y(\mathfrak{gl}(m|n))$ and $Y(\mathfrak{gl}(n|m))$
\be{morph-dBV}
\varphi\Bigl(\mathbb{C}^{m|n}(\ort)\Bigr)=
\frac{\mathbb{C}^{n|m}(\olt)}
{\prod_{k=1}^N\alpha_{N+1-k}(\bt^k)}.
\ee

\section{Main results\label{S-MR}}

In this section we present the main results of the paper.
They are of three types:  recursion formulas for Bethe vectors;  sum formula for the Bethe vectors scalar product;
recursion formulas for the scalar product highest coefficients. Recall that we formally consider the case $m,n\ne 0$. However,
in subsection~\ref{SS-SEglm} we specify our results to the particular case of $\mathfrak{gl}(m)$-based models, that is, $n=0$.
The case $m=0$ can be obtained from the latter via replacement $c\to -c$ in the $R$-matrix \eqref{R-mat}.

\subsection{Recursion for Bethe vectors\label{SS-RfBV}}

Here we give recursions for (dual) Bethe vectors. These recursions allow us to construct Bethe vectors, knowing the ones depending
on a smaller number of parameters. The corresponding proofs are given in section~\ref{S-RfBV}.

\begin{prop}\label{P-recBV}
Bethe vectors of $\mathfrak{gl}(m|n)$-based models satisfy a recursion
\begin{multline}\label{recBV-G01}
\mathbb{B}(\bigr\{z,\bt^1\bigr\};\bigl\{\bt^k\bigr\}_{2}^{N})=
\sum_{j=2}^{N+1}\frac{T_{1,j}(z)}{\lambda_2(z)}
\sum_{\text{\rm part}(\bt^2,\dots,\bt^{j-1})}\mathbb{B}(\bigr\{\bt^1\bigr\};\bigl\{\bt^k_{\st}\bigr\}_{2}^{j-1};\bigl\{\bt^k\bigr\}_{j}^{N})\\
\times
\frac{\prod_{\nu=2}^{j-1}\alpha_\nu(\bt^\nu_{\so})g_{[\nu]}(\bt^{\nu}_{\so},\bt^{\nu-1}_{\so})
\gamma_{\nu}(\bt^{\nu}_{\st},\bt^{\nu}_{\so})}{h(\bt^1,z)^{\delta_{m,1}}\prod_{\nu=1}^{j-1}f_{[\nu+1]}(\bt^{\nu+1},\bt^{\nu}_{\so})}.
\end{multline}
Here for $j>2$ the sets of Bethe parameters $\bt^2,\dots,\bt^{j-1}$ are divided into disjoint subsets $\bt^\nu_{\so}$ and
$\bt^\nu_{\st}$ ($\nu=2,\dots,j-1$) such that the subset $\bt^\nu_{\so}$ consists of one element only: $\#\bt^\nu_{\so}=1$.
The sum is taken over all partitions of this type. We
set by definition $\bt^1_{\so}\equiv z$ and $\bt^{N+1}=\emptyset$.
\end{prop}

We used the following notation in proposition~\ref{P-recBV}
\begin{equation}\label{BV-partsets}
\begin{aligned}
&\mathbb{B}(\bigr\{z,\bt^1\bigr\};\bigl\{\bt^k\bigr\}_{2}^{N})=\mathbb{B}(\bigr\{z,\bt^1\bigr\};\bt^2;\dots;\bt^{N}),\\
&\mathbb{B}(\bigr\{\bt^1\bigr\};\bigl\{\bt^k_{\st}\bigr\}_{2}^{j-1};\bigl\{\bt^k\bigr\}_{j}^{N})
=\mathbb{B}(\bt^1;\bt^2_{\st};\dots;\bt^{j-1}_{\st};\bt^j;\dots;\bt^{N}).
\end{aligned}
\end{equation}
This and similar notation will be used throughout of the paper.

{\sl Remark.} We stress that each of the subsets $\bt^2_{\so},\dots,\bt^{N}_{\so}$ in \eqref{recBV-G01} must consist of exactly one element.
However, this condition is not feasible, if the original Bethe vector $\mathbb{B}(t)$ contains an empty set $\bt^k=\emptyset$ for some
$k\in[2,\dots,N]$. In this case, the sum over $j$ in \eqref{recBV-G01} breaks off at $j=k$.  Indeed, the action of the operators
$T_{1,j}(z)$ with $j>k$ on a Bethe vector necessarily creates a quasiparticle of the color $k$. Since this quasiparticle is absent in the lhs
of \eqref{recBV-G01}, we cannot have the operators $T_{1,j}(z)$ with $j>k$ in the rhs. Similar consideration shows that
if $\mathbb{B}(t)$ contains several empty sets
$\bt^{k_1},\dots,\bt^{k_\ell}$, then the sum ends at $j=\min(k_1,\dots,k_\ell)$.

{\sl Remark.}  One can notice that for $m=1$  an additional factor $h(\bt^1,z)^{-1}$ appears in the recursion.
The point is that with this recursion we add a quasiparticle of the color $1$ to the original set of quasiparticles
via the actions of the operators $T_{1,j}$. For $m=1$ all these operators are odd, which explains the appearance of the factor
$h(\bt^1,z)^{-1}$. This difference can also be seen explicitly in the example of recursion for the main term \eqref{Norm-MT}
\begin{equation}\label{recMT}
\widetilde{\mathbb{B}}(\bigr\{z,\bt^1\bigr\};\bigl\{\bt^k\bigr\}_{2}^{N})=
\frac{T_{1,2}(z)\widetilde{\mathbb{B}}(\bt)}{h(\bt^1,z)^{\delta_{m,1}}\lambda_2(z)  f_{[2]}(\bt^{2},z)}\; .
\end{equation}

Using the mappings \eqref{morphism} and \eqref{antimo} one can obtain one more recursion for the Bethe vectors and
two recursions for the dual ones.

\begin{prop}\label{C2-recBV}
Bethe vectors of $\mathfrak{gl}(m|n)$-based models satisfy a recursion
\begin{multline}\label{recBV-G02pr}
\mathbb{B}(\bigl\{\bt^k\bigr\}_{1}^{N-1};\bigr\{z,\bt^N\bigr\})=
\sum_{j=1}^{N} \frac{T_{j,N+1}(z)}{\lambda_{N+1}(z)}
\sum_{\text{\rm part}(\bt^j,\dots,\bt^{N-1})}\mathbb{B}(\bigr\{\bt^k\bigr\}_1^{j-1};\bigl\{\bt^k_{\st}\bigr\}_{j}^{N-1};\bt^{N})\\
\times
\frac{\prod_{\nu=j}^{N-1}g_{[\nu+1]}(\bt^{\nu+1}_{\so},\bt^{\nu}_{\so})
\hg_{\nu}(\bt^{\nu}_{\so},\bt^{\nu}_{\st})}{h(\bt^N,z)^{\delta_{m,N}}\prod_{\nu=j}^{N}f_{[\nu]}(\bt^{ \nu}_{\so},\bt^{\nu-1})}.
\end{multline}
Here for $j<N$ the sets of Bethe parameters $\bt^j,\dots,\bt^{N-1}$ are divided into disjoint subsets $\bt^\nu_{\so}$ and
$\bt^\nu_{\st}$ ($\nu=j,\dots,N-1$) such that the subset $\bt^\nu_{\so}$ consists of one element: $\#\bt^\nu_{\so}=1$.
The sum is taken over all partitions of this type. We
set by definition $\bt^N_{\so}\equiv z$ and $\bt^{0}=\emptyset$.
\end{prop}

{\sl Remark.} If the Bethe vector $\mathbb{B}(t)$ contains several empty sets
$\bt^{k_1},\dots,\bt^{k_\ell}$, then the sum  over $j$ in \eqref{recBV-G02pr} begins with $j=\max(k_1,\dots,k_\ell)+1$.

Acting  with antimorphism \eqref{antimo} onto equations \eqref{recBV-G01} and \eqref{recBV-G02pr} we immediately arrive at recursions for the
dual Bethe vectors.


\begin{cor}\label{C-recDBV}
Dual Bethe vectors of $\mathfrak{gl}(m|n)$-based models satisfy recursions
\begin{multline}\label{recdBV-G01}
\mathbb{C}(\bigr\{z,\bs^1\bigr\};\bigl\{\bs^k\bigr\}_{2}^{N})=
\sum_{j=2}^{N+1}
\sum_{\text{\rm part}(\bs^2,\dots,\bs^{j-1})}\mathbb{C}(\bigr\{\bs^1\bigr\};\bigl\{\bs^k_{\st}\bigr\}_{2}^{j-1};\bigl\{\bs^k\bigr\}_{j}^{N})
\frac{T_{j,1}(z)}{\lambda_2(z)} (-1)^{r_1\delta_{m,1}} \\
\times
\frac{\prod_{\nu=2}^{j-1}\alpha_\nu(\bs^\nu_{\so})g_{[\nu]}(\bs^{\nu}_{\so},\bs^{\nu-1}_{\so})
\hg_{\nu}(\bs^{\nu}_{\st},\bs^{\nu}_{\so})}{h(\bs^1,z)^{\delta_{m,1}}\prod_{\nu=1}^{j-1}f_{[\nu+1]}(\bs^{\nu+1},\bs^{\nu}_{\so})},
\end{multline}
and
\begin{multline}\label{C-recDBV2}
\mathbb{C}(\bigl\{\bs^k\bigr\}_{1}^{N-1};\bigr\{z,\bs^N\bigr\})=
\sum_{j=1}^{N}
\sum_{\text{\rm part}(\bs^j,\dots,\bs^{N-1})}\mathbb{C}(\bigr\{\bs^k\bigr\}_1^{j-1};\bigl\{\bs^k_{\st}\bigr\}_{j}^{N-1};\bs^{N})
\frac{T_{N+1,j}(z)}{\lambda_{N+1}(z)} (-1)^{r_N\delta_{m,N}}  \\
\times
\frac{\prod_{\nu=j}^{N-1}g_{[\nu]}(\bs^{\nu+1}_{\so},\bs^{\nu}_{\so})
\gamma_{\nu}(\bs^{\nu}_{\so},\bs^{\nu}_{\st})}{h(\bs^N,z)^{\delta_{m,N}}\prod_{\nu=j}^{N}f_{[\nu]}(\bs^{ \nu}_{\so},\bs^{\nu-1})}.
\end{multline}
Here the summation over the partitions occurs as in the formulas \eqref{recBV-G01} and \eqref{recBV-G02pr}. The numbers $r_1$ (resp. $r_N$)
are the cardinalities of the sets $\bs^1$  (resp. $\bs^N$).
The subsets $\bs^\nu_{\so}$ consist of one element: $\#\bs^\nu_{\so}=1$. If $\mathbb{C}(\bs)$ contains empty sets
of the Bethe parameters, then the sums cut similarly to the case of the Bethe vectors $\mathbb{B}(\bt)$.
By definition $\bs^1_{\so}\equiv z$ in \eqref{recdBV-G01},  $\bs^N_{\so}\equiv z$ in \eqref{C-recDBV2}, and $\bs^{0}=\bs^{N+1}=\emptyset$.
\end{cor}

The proof of corollary~\ref{C-recDBV} is given in section~\ref{SS-RfdBV}.

Using recursion \eqref{recBV-G01} one can express a Bethe vector with $\#\bt^1=r_1$ in terms of Bethe vectors with $\#\bt^1=r_1-1$. Applying
this recursion successively we eventually express the original Bethe vector in terms of a linear combination of
terms that are products of the monodromy matrix
elements $T_{1,j}$ acting onto  Bethe vectors with
$\#\bt^1=0$. The latter effectively corresponds to the Yangian $Y(\mathfrak{gl}(m-1|n))$ (see \cite{HutLPRS17a}):
\be{BVmnBVm1n}
\mathbb{B}^{m|n}(\emptyset;\{\bt^k\}_2^N)=\mathbb{B}^{m-1|n}(\bt)\Bigr|_{\bt^k\to \bt^{k+1}}.
\ee
Thus, continuing this process we formally can reduce Bethe vectors of $Y(\mathfrak{gl}(m|n))$ to the ones of $Y(\mathfrak{gl}(1|n))$.

Similarly, using recursion \eqref{recBV-G02pr} and
\be{BVmnBVmn1}
\mathbb{B}^{m|n}(\{\bt^k\}_1^{N-1};\emptyset)=\mathbb{B}^{m|n-1}(\bt),
\ee
we eventually reduce Bethe vectors of $Y(\mathfrak{gl}(m|n))$ to the ones of $Y(\mathfrak{gl}(m|1))$. The combination of both recursions thus defines
a unique procedure for constructing Bethe vectors with respect to the known Bethe vectors of $Y(\mathfrak{gl}(1|1))$:
$\mathbb{B}^{1|1}(\bt)=\bT_{1,2}(\bt)|0\rangle/\lambda_2(\bt)$. Similarly, one can built dual Bethe vectors via \eqref{recdBV-G01}, \eqref{C-recDBV2}.
These procedures, of course, are of little use for practical purposes, however, they can be used to prove various assertions by induction.

\subsection{Sum formula for the scalar product\label{SS-SFSP}}

Let $\mathbb{B}(\bt)$ be a generic Bethe vector and   $\mathbb{C}(\bs)$ be a generic dual Bethe vector such that
$\#\bt^k=\#\bs^k=r_k$, $k=1,\dots,N$. Then their
scalar product is defined by
\be{SP-def}
S(\bs|\bt)= \mathbb{C}(\bs) \mathbb{B}(\bt).
\ee
Note that if $\#\bt^k\ne\#\bs^k$ for some  $k\in\{1,\dots,N\}$, then the scalar product vanishes. Indeed, in this case the numbers
of creation and annihilation operators of the color $k$ do not coincide.

Applying \eqref{Avac} to the scalar product and using $\bigl[\mathbb{B}(\bt)\bigr]=\bigl[\mathbb{C}(\bt)\bigr]=r_m$ \cite{HutLPRS17a} we
find that
\be{SP-def1}
S(\bs|\bt)= \mathbb{C}(\bt) \mathbb{B}(\bs)=S(\bt|\bs).
\ee

Computing the scalar product one should use commutation relations \eqref{TM-1} and move all operators $T_{i,j}$ with $i>j$
from the dual vector $\mathbb{C}(\bs)$ to the right through the  operators $T_{i,j}$ with $i<j$, which are in the vector $\mathbb{B}(\bt)$.
In the process of commutation, new operators will appear, which should be moved to the right or left, depending on the relation between their subscripts.
Once an operator $T_{i,j}$ with $i\ge j$ reaches the vector $|0\rangle$, it either annihilates it for $i> j$, or gives a function $\lambda_i$
for $i=j$. The argument of the function $\lambda_i$ can a priori be any Bethe parameter $t^k_\ell$ or $s^k_\ell$. Similarly,
if an operator $T_{i,j}$ with $i\le j$ reaches the vector $\langle0|$, it either annihilates it for $i<j$, or gives a function $\lambda_i$
for $i=j$, which depends on one of the Bethe parameters.

Due to the normalization of the Bethe vectors the functions $\lambda_i$ then turn into the ratios $\alpha_i$. Thus,
the scalar product eventually depends on the functions $\alpha_i$ and some rational functions which appear
in the process of commutating the monodromy matrix entries.

The following proposition specifies how the scalar product depends on the functions $\alpha_i$.

\begin{prop}\label{P-al-dep}
Let $\mathbb{B}(\bt)$ be a generic Bethe vector and   $\mathbb{C}(\bs)$ be a generic dual Bethe vector such that
$\#\bt^k=\#\bs^k=r_k$, $k=1,\dots,N$. Then their
scalar product is given by
\begin{equation}\label{al-dep}
S(\bs|\bt)= \sum W^{m|n}_{\text{\rm part}}(\bs_{\so},\bs_{\st}|\bt_{\so},\bt_{\st})
  \prod_{k=1}^{N} \alpha_{k}(\bs^k_{\so}) \alpha_{k}(\bt^k_{\st}).
\end{equation}
Here all the sets of the Bethe parameters $\bt^k$ and $\bs^k$ are divided into two subsets $\bt^k\Rightarrow\{\bt^k_{\so}, \bt^k_{\st} \}$
and $\bs^k\Rightarrow\{\bs^k_{\so}, \bs^k_{\st} \}$, such that $\#\bt^k_{\so}=\#\bs^k_{\so}$. The sum is taken over all possible partitions of this
type.
The rational coefficients $W^{m|n}_{\text{\rm part}}$ depend on the partition. They are completely determined by the
$R$-matrix of the model and do not depend on the ratios of the vacuum eigenvalues $\alpha_k$.
\end{prop}

Proposition~\ref{P-al-dep} states that after calculating the scalar product the Bethe parameters of the type $k$ ($t^k_j$ or $s^k_j$) can be arguments
of functions  $\lambda_{k+1}$ or $\lambda_k$ only.  Due to the normalization  of the Bethe vectors  these functions  respectively cancel in the first case
or produce the functions $\alpha_k$ in the second case. We prove proposition~\ref{P-al-dep} in section~\ref{SS-HSCPD}.

We would like to stress that the rational functions $W^{m|n}_{\text{\rm part}}$ are model independent. Indeed, within the QISM framework the Hamiltonian of a quantum model is encoded in the supertrace of the monodromy matrix $T(u)$. Thus, one can say that the quantum model is defined by $T(u)$. Looking
at presentation \eqref{al-dep} one can notice that the model dependent part of the scalar product entirely lies in the $\alpha_k$ functions, because only
these functional parameters depend on the monodromy matrix. On the other hand, the coefficients $W^{m|n}_{\text{\rm part}}$ are completely determined by the
$R$-matrix, that is, they depend only on the underlying algebra. Thus, if two different quantum integrable models have the same $R$-matrix \eqref{R-mat}, then the scalar products of Bethe vectors in these models are given by  \eqref{al-dep} with the same coefficients $W^{m|n}_{\text{\rm part}}$.

The Highest Coefficient (HC) of the scalar product is defined as a rational coefficient corresponding to the partition $\bs_{\so}=\bs$, $\bt_{\so}=\bt$,
and  $\bs_{\st}=\bt_{\st}=\emptyset$.
We denote the HC by $Z^{m|n}(\bs|\bt)$. Then,
the HC is a particular case of the rational coefficient\footnote{Note that we have changed the definition of the HC with respect to
the one that we used in our previous publications. Now it involves a normalization factor
$\prod_{j=1}^{N-1} f_{[j+1]}(\bar s^{j+1},\bar s^j) f_{[j+1]}(\bar t^{j+1},\bar t^j)$.}
$W^{m|n}_{\text{\rm part}}$:
\be{HCdef}
W^{m|n}_{\text{\rm part}}(\bs,\emptyset|\bt,\emptyset)=Z^{m|n}(\bs|\bt).
\ee
Similarly one can  define a conjugated HC $\overline{Z}^{m|n}(\bs|\bt)$ as a coefficient corresponding to the partition
 $\bs_{\st}=\bs$, $\bt_{\st}=\bt$, and $\bs_{\so}=\bt_{\so}=\emptyset$.
\be{cHCdef}
W^{m|n}_{\text{\rm part}}(\emptyset,\bs|\emptyset,\bt)=\overline{Z}^{m|n}(\bs|\bt).
\ee
Due to \eqref{SP-def1} one can easily show that
\be{Z-bZ}
\overline{Z}^{m|n}(\bs|\bt)=Z^{m|n}(\bt|\bs).
\ee

The following proposition determines the general coefficient $W^{m|n}_{\text{\rm part}}$ in terms of the HC.

\begin{prop}\label{P-Hyp-Resh}
For a fixed partition $\bt^k\Rightarrow\{\bt^k_{\so}, \bt^k_{\st} \}$
and $\bs^k\Rightarrow\{\bs^k_{\so}, \bs^k_{\st} \}$ in \eqref{al-dep} the rational coefficient
 $W^{m|n}_{\text{\rm part}}$ has the following presentation in terms of the HC:
\begin{equation}\label{HypResh}
W^{m|n}_{\text{\rm part}}(\bs_{\so},\bs_{\st}|\bt_{\so},\bt_{\st}) =  Z^{m|n}(\bs_{\so}|\bt_{\so}) \;\; Z^{m|n}(\bt_{\st}|\bs_{\st})
\;  \frac{\prod_ {k=1}^{N}  \gamma_{k}(\bs^k_{\st},\bs^k_{\so}) \gamma_{k}(\bt^k_{\so},\bt^k_{\st})}
{\prod_{j=1}^{N-1} f_{[j+1]}(\bar s^{j+1}_{\st},\bar s^j_{\so})
 f_{[j+1]}(\bar t^{j+1}_{\so},\bar t^j_{\st})}.
\end{equation}
\end{prop}

The proof of proposition~\ref{P-Hyp-Resh} is given in section~\ref{S-RF}.

Explicit expressions for the HC are known for small $m$ and $n$ \cite{HutLPRS16c}. In particular,
\be{Z11}
 Z^{1|1}(\bs|\bt)=g(\bs,\bt).
 \ee
Determinant representations for $Z^{2|0}$ or $Z^{0|2}$ were obtained in \cite{Ize87}. Relatively compact formulas for $Z^{m|n}$
at $m+n=3$ were found in \cite{Whe12,BelPRS12b,HutLPRS16c}, however,
representations for the HC in the general $\mathfrak{gl}(m|n)$ case are very cumbersome. Instead, one can use relatively simple
recursions established by the following propositions.

\begin{prop}\label{P-1HCrec}
The HC $Z^{m|n}(\bs|\bt)$ possesses the following recursion over the set $\bs^1$:
\begin{multline}\label{Rec-HC1}
Z^{m|n}(\bs|\bt)=\sum_{p=2}^{N+1}
\sum_{\substack{\text{\rm part}(\bs^2,\dots,\bs^{p-1})\\
\text{\rm part}(\bt^1,\dots,\bt^{p-1})}}
\frac{g_{[2]}(\bt^{1}_{\so},\bs^{1}_{\so})\gamma_{1}(\bt^{1}_{\so},\bt^{1}_{\st})
f(\bt^{1}_{\st}, \bs^{1}_{\so} )}{f_{[p]}(\bs^{p},\bs^{p-1}_{\so})h(\bs^1,\bs^1_{\so})^{\delta_{m,1}}}\\
\times \prod_{\nu=2}^{p-1}\frac{g_{[\nu]}(\bs^{\nu}_{\so},\bs^{\nu-1}_{\so})g_{[\nu+1]}(\bt^{\nu}_{\so},\bt^{\nu-1}_{\so})
\gamma_{\nu}(\bs^{\nu}_{\st},\bs^{\nu}_{\so})    \gamma_{\nu}(\bt^{\nu}_{\so},\bt^{\nu}_{\st})}
{ f_{[\nu]}(\bs^{\nu},\bs^{\nu-1}_{\so})f_{[\nu]}(\bar t^{\nu}_{\so},\bar t^{\nu-1})}\\
\times Z^{m|n}(\bigl\{\bs^k_{\st}\bigr\}_1^{p-1},\bigl\{\bs^{k}\bigr\}_p^N|
\bigl\{\bt^k_{\st}\bigr\}_1^{p-1};\bigl\{\bt^k\bigr\}_p^{N}).
\end{multline}

Here for every  fixed $p\in\{2,\dots,m+n\}$ the sums are taken over partitions $\bt^k\Rightarrow\{\bt^k_{\so}, \bt^k_{\st} \}$
with $k=1,\dots,p-1$
and $\bs^k\Rightarrow\{\bs^k_{\so}, \bs^k_{\st} \}$ with $k=2,\dots,p-1$, such that $\#\bt^k_{\so}=\#\bs^k_{\so}=1$
 for $k=2,...,p-1$.
The subset $\bs^1_{\so}$
is a fixed Bethe parameter from the set $\bs^1$. There is no sum over partitions of the set $\bs^1$ in \eqref{Rec-HC1}.
\end{prop}

The proof of this proposition is given in section~\ref{SS-PRHC}.

\begin{cor}\label{P-2HCrec}
The HC $Z^{m|n}(\bs|\bt)$ satisfies the following recursion over the set $\bt^N$:
\begin{multline}\label{Rec-2HC1}
Z^{m|n}(\bs|\bt)=\sum_{p=1}^{N}
\sum_{\substack{\text{\rm part}(\bs^p,\dots,\bs^{N})\\\text{\rm part}(\bt^p,\dots,\bt^{N-1})}}
\frac{g(\bs^{N}_{\so},\bt^{N}_{\so})\hg_{N}(\bs^{N}_{\st},\bs^{N}_{\so})
f(\bs^{N}_{\st}, \bt^{N}_{\so})}{f_{[p]}(\bt^{p}_{\so},\bt^{p-1})h(\bt^N,\bt^N_{\so})^{\delta_{m,N}}}
\\
\times \prod_{\nu=p}^{N-1}\frac{g_{[\nu+1]}(\bs^{\nu+1}_{\so},\bs^{\nu}_{\so})g_{[\nu+1]}(\bt^{\nu+1}_{\so},\bt^{\nu}_{\so})
\hg_{\nu}(\bs^{\nu}_{\st},\bs^{\nu}_{\so})\hg_{\nu}(\bt^{\nu}_{\so},\bt^{\nu}_{\st})}
{f_{[\nu+1]}(\bs^{\nu+1},\bs^{\nu}_{\so})f_{[\nu+1]}(\bar t^{\nu+1}_{\so},\bar t^{\nu})}\\
\times Z^{m|n}(\bigl\{\bs^k\bigr\}_1^{p-1},\bigl\{\bs^{k}_{\st}\bigr\}_p^N|
\bigl\{\bt^k\bigr\}_1^{p-1};\bigl\{\bt^k_{\st}\bigr\}_p^{N}).
\end{multline}
Here for every  fixed $p\in\{1,\dots,m+n-1\}$ the sums are taken over partitions $\bt^k\Rightarrow\{\bt^k_{\so}, \bt^k_{\st} \}$
with $k=p,\dots,N-1$
and $\bs^k\Rightarrow\{\bs^k_{\so}, \bs^k_{\st} \}$ with $k=p,\dots,N$, such that $\#\bt^k_{\so}=\#\bs^k_{\so}=1$
 for $k=p,\dots,N-1$.
The subset $\bt^N_{\so}$
is a fixed Bethe parameter from the set $\bt^N$. There is no sum over partitions of the set $\bt^N$ in \eqref{Rec-2HC1}.
\end{cor}

This recursion follows from \eqref{Rec-HC1} and a symmetry property of the HC \eqref{Z-Z} proved in section~\ref{SS-SHC}.

{\sl Remark.} Similarly to the recursions for the Bethe vectors the sums over $p$ in \eqref{Rec-HC1}, \eqref{Rec-2HC1}
break off, if HC $Z^{m|n}(\bs|\bt)$ contains empty sets of the Bethe parameters. If the colors of the empty sets are $\{k_1,\dots,k_\ell\}$,
then the sum over  $p$ ends at $p=\min(k_1,\dots,k_\ell)$ in the recursion \eqref{Rec-HC1}, while in the recursion \eqref{Rec-2HC1} it
begins at $p =\max(k_1,\dots,k_\ell)+1$ . These restrictions follow from the corresponding restrictions in the recursions for the Bethe vectors.

Using proposition~\ref{P-1HCrec} one can built the HC with $\#\bs^1=\#\bt^1=r_1$ in terms of the HC with $\#\bs^1=\#\bt^1=r_1-1$. In particular,
$Z^{m|n}$ with $\#\bs^1=\#\bt^1=1$ can be expressed in terms of $Z^{m|n}$ with $\#\bs^1=\#\bt^1=0$.  It is obvious, however, that
\be{ZmnZm-1n}
Z^{m|n}(\emptyset,\{\bs^k\}_2^N|\emptyset,\{\bt^k\}_2^N)=Z^{m-1|n}(\{\bs^k\}_2^N|\{\bt^k\}_2^N).
\ee
due to \eqref{BVmnBVm1n}.
Thus, equation \eqref{Rec-HC1} allows one to perform recursion over $m$ as well.

Similarly, corollary~\ref{P-2HCrec} allows one to find the HC with $\#\bs^N=\#\bt^N=r_N$ in terms of the HC with $\#\bs^N=\#\bt^N=r_N-1$ and
to perform recursion over $n$.

Thus, using recursions \eqref{Rec-HC1} and \eqref{Rec-2HC1} one can eventually express $Z^{m|n}(\bs|\bt)$ in terms of known HC, say,
for $m+n=2$. However, the corresponding explicit expressions hardly can be used in practice, because they are too bulky.
At the same time, these recursions appear be very useful for proofs of some important properties of HC.

\subsection{Simplified expressions for models with $\mathfrak{gl}(m)$ symmetry\label{SS-SEglm}}

As already mentioned, the results stated above are also valid for the case of $\mathfrak{gl}(m)$ Lie algebras with $m>1$, simply by setting $n=0$.
This implies $N=m-1$.
In that case, most of expressions simplify, due to the absence of grading. We present here the simplified results occurring for $\mathfrak{gl}(m)$.
\begin{itemize}
\item Bethe vectors of $\mathfrak{gl}(m)$-based models satisfy the recursions
\begin{multline}
\mathbb{B}(\bigr\{z,\bt^1\bigr\};\bigl\{\bt^k\bigr\}_{2}^{m-1})=
\sum_{j=2}^{m}\frac{T_{1,j}(z)}{\lambda_2(z)}
\sum_{\text{\rm part}(\bt^2,\dots,\bt^{j-1})}\mathbb{B}(\bigr\{\bt^1\bigr\};\bigl\{\bt^k_{\st}\bigr\}_{2}^{j-1};\bigl\{\bt^k\bigr\}_{j}^{m-1})\\
\times
\frac{\prod_{\nu=2}^{j-1}\alpha_\nu(\bt^\nu_{\so})\,g(\bt^{\nu}_{\so},\bt^{\nu-1}_{\so})
f(\bt^{\nu}_{\st},\bt^{\nu}_{\so})} {\prod_{\nu=1}^{j-1}f(\bt^{\nu+1},\bt^{\nu}_{\so})},
\end{multline}
where the conditions on sets of Bethe parameters are the same as in Proposition \ref{P-recBV},
\begin{multline}
\mathbb{B}(\bigl\{\bt^k\bigr\}_{1}^{m-2};\bigr\{z,\bt^{m-1}\bigr\})=
\sum_{j=1}^{m-1} \frac{T_{j,m}(z)}{\lambda_{m}(z)}
\sum_{\text{\rm part}(\bt^j,\dots,\bt^{m-2})}\mathbb{B}(\bigr\{\bt^k\bigr\}_1^{j-1};\bigl\{\bt^k_{\st}\bigr\}_{j}^{m-2};\bt^{m-1})\\
\times
\frac{\prod_{\nu=j}^{m-2}g(\bt^{\nu+1}_{\so},\bt^{\nu}_{\so})
f(\bt^{\nu}_{\so},\bt^{\nu}_{\st})}{\prod_{\nu=j}^{m-1}f(\bt^{ \nu}_{\so},\bt^{\nu-1})},
\end{multline}
where the conditions on sets of Bethe parameters are the same as in Proposition \ref{C2-recBV}.
The starting point for these recursions is the $\mathfrak{gl}(2)$ Bethe vector $\mathbb{B}(\bt)=T_{12}(\bt)|0\rangle /\lambda_2(\bt)$.
\item
Dual Bethe vectors of $\mathfrak{gl}(m)$-based models satisfy the recursions
\begin{multline}
\mathbb{C}(\bigr\{z,\bs^1\bigr\};\bigl\{\bs^k\bigr\}_{2}^{m-1})=
\sum_{j=2}^{m}
\sum_{\text{\rm part}(\bs^2,\dots,\bs^{j-1})}\mathbb{C}(\bigr\{\bs^1\bigr\};\bigl\{\bs^k_{\st}\bigr\}_{2}^{j-1};\bigl\{\bs^k\bigr\}_{j}^{m-1})
\frac{T_{j,1}(z)}{\lambda_2(z)}\\
\times
\frac{\prod_{\nu=2}^{j-1}\alpha_\nu(\bs^\nu_{\so})g(\bs^{\nu}_{\so},\bs^{\nu-1}_{\so})
f(\bs^{\nu}_{\st},\bs^{\nu}_{\so})}{\prod_{\nu=1}^{j-1}f(\bs^{\nu+1},\bs^{\nu}_{\so})},
\end{multline}
and
\begin{multline}\mathbb{C}(\bigl\{\bs^k\bigr\}_{1}^{m-2};\bigr\{z,\bs^{m-1}\bigr\})=
\sum_{j=1}^{m-1}
\sum_{\text{\rm part}(\bs^j,\dots,\bs^{m-2})}\mathbb{C}(\bigr\{\bs^k\bigr\}_1^{j-1};\bigl\{\bs^k_{\st}\bigr\}_{j}^{m-2};\bs^{m-1})
\frac{T_{m,j}(z)}{\lambda_{m}(z)}\\
\times
\frac{\prod_{\nu=j}^{m-2}g(\bs^{\nu+1}_{\so},\bs^{\nu}_{\so})
f(\bs^{\nu}_{\so},\bs^{\nu}_{\st})}{\prod_{\nu=j}^{m-1}f(\bs^{ \nu}_{\so},\bs^{\nu-1})}.
\end{multline}
The conditions on the sets of parameters and partitions are given in Corollary \ref{C-recDBV}.
The starting point for these recursions is the $\mathfrak{gl}(2)$ dual Bethe vector $\mathbb{C}(\bt)=\langle0|T_{21}(\bt) / \lambda_2(\bt)$.
\item For a fixed partition $\bt^k\Rightarrow\{\bt^k_{\so}, \bt^k_{\st} \}$
and $\bs^k\Rightarrow\{\bs^k_{\so}, \bs^k_{\st} \}$ in \eqref{al-dep} the rational coefficient
 $W^{m}_{\text{\rm part}}$ has the following presentation in terms of the HC:
\begin{equation}\label{Resh-gln}
W^{m}_{\text{\rm part}}(\bs_{\so},\bs_{\st}|\bt_{\so},\bt_{\st}) =  Z^{m}(\bs_{\so}|\bt_{\so}) \; Z^{m}(\bt_{\st}|\bs_{\st})
\;  \frac{\prod_ {k=1}^{m-1}  f(\bs^k_{\st},\bs^k_{\so}) f(\bt^k_{\so},\bt^k_{\st})}
{\prod_{j=1}^{m-2} f(\bar s^{j+1}_{\st},\bar s^j_{\so})
 f(\bar t^{j+1}_{\so},\bar t^j_{\st})}.
\end{equation}

In the $\mathfrak{gl}(2)$ and $\mathfrak{gl}(3)$ cases this expression reduces to the formulas respectively obtained in  \cite{Kor82} and \cite{Res86}.

\item The HC $Z^{m}(\bs|\bt)$ possesses the following recursions:
\begin{multline}
Z^{m}(\bs|\bt)=\sum_{p=2}^{m}
\sum_{\substack{\text{\rm part}(\bs^2,\dots,\bs^{p-1})\\
\text{\rm part}(\bt^1,\dots,\bt^{p-1})}}
\frac{g(\bt^{1}_{\so},\bs^{1}_{\so})f(\bt^{1}_{\so},\bt^{1}_{\st})
f(\bt^{1}_{\st},\bs^{1}_{\so})}{f(\bs^{p},\bs^{p-1}_{\so})}\\
\times \prod_{\nu=2}^{p-1}\frac{g(\bs^{\nu}_{\so},\bs^{\nu-1}_{\so})g(\bt^{\nu}_{\so},\bt^{\nu-1}_{\so})
f(\bs^{\nu}_{\st},\bs^{\nu}_{\so}) f(\bt^{\nu}_{\so},\bt^{\nu}_{\st})}
{ f(\bs^{\nu},\bs^{\nu-1}_{\so})f(\bar t^{\nu}_{\so},\bar t^{\nu-1})}\\
\times Z^{m}(\bigl\{\bs^k_{\st}\bigr\}_1^{p-1},\bigl\{\bs^{k}\bigr\}_p^{m-1}|
\bigl\{\bt^k_{\st}\bigr\}_1^{p-1};\bigl\{\bt^k\bigr\}_p^{m-1}),
\end{multline}
and
\begin{multline}
Z^{m}(\bs|\bt)=\sum_{p=1}^{m-1}
\sum_{\substack{\text{\rm part}(\bs^p,\dots,\bs^{m-1})\\\text{\rm part}(\bt^p,\dots,\bt^{m-2})}}
\frac{  g(\bt^{m-1}_{\so},\bs^{m-1}_{\so}) f(\bs^{m-1}_{\st},\bs^{m-1}_{\so})
f(\bt^{m-1}_{\so},\bs^{m-1}_{\st})  }{f(\bt^{p}_{\so},\bt^{p-1})}
\\
\times \prod_{\nu=p}^{m-2}\frac{g(\bs^{\nu+1}_{\so},\bs^{\nu}_{\so})g(\bt^{\nu+1}_{\so},\bt^{\nu}_{\so})
f(\bs^{\nu}_{\st},\bs^{\nu}_{\so}) f(\bt^{\nu}_{\so},\bt^{\nu}_{\st})}
{f(\bs^{\nu+1},\bs^{\nu}_{\so}) f(\bar t^{\nu+1}_{\so},\bar t^{\nu})}\\
\times Z^{m}(\bigl\{\bs^k\bigr\}_1^{p-1},\bigl\{\bs^{k}_{\st}\bigr\}_p^{m-1}|
\bigl\{\bt^k\bigr\}_1^{p-1};\bigl\{\bt^k_{\st}\bigr\}_p^{m-1}).
\end{multline}
The conditions on the sets of parameters and partitions are given in Proposition \ref{P-1HCrec} and Corollary \ref{P-2HCrec}.
Here, the starting point corresponds to the $\mathfrak{gl}(2)$ case,  in which  $Z^{2}(\bs|\bt)$ is equal to the
partition function of the six-vertex model with domain wall boundary conditions \cite{Kor82,Ize87}.
\end{itemize}

\section{Proof of recursion for Bethe vectors\label{S-RfBV}}

 One can prove  proposition~\ref{P-recBV} via the formulas of the operators $T_{1,j}(z)$ action onto the Bethe vector. These formulas were derived
in \cite{HutLPRS17a}
\begin{multline}\label{T1jactBV}
T_{1,j}(z)\mathbb{B}(\bt)= \eta_j\mathbb{B}(\{z,\bt^k\}_1^{j-1};\{\bt^k\}_j^{N})\\
+\sum_{q=j+1}^{N+1}\sum_{\text{part}(\bt^j,\dots,\bt^{q-1})}H_{q,j}(\text{part})
\mathbb{B}(\{z,\bt^k\}_1^{j-1};\{z,\bt^k_{\st}\}_j^{q-1};\{\bt^k\}_q^{N}).
\end{multline}
Here in the second line for every $q$ we have a sum over partitions of the sets $\bt^j,\dots,\bt^{q-1}$.
The coefficient $\eta_j$ in \eqref{T1jactBV} is
\be{alj}
\eta_j=\lambda_j(z)f_{[j]}(\bt^j,z)h(\bt^m,z)^{[j]}.
\ee
The coefficient $H_{q,j}$ depends on the partitions and has the form
\begin{equation}\label{Hqj}
H_{q,j}(\text{part})=f_{[q]}(\bt^q,z)h(\bt^m,z)^{[j]}h(\bt^m_{\st},z)^{[q]-[j]} \lambda_q(z)
g_{[j]}(z,\bt_{\so}^{q-1})\prod_{\nu=j+1}^{q-1}g_{[\nu]}(\bt^{\nu}_{\so},\bt^{\nu-1}_{\so})
\prod_{\nu=j}^{q-1}\Omega_\nu,
\end{equation}
where
\be{Om}
\Omega_\nu=
\frac{\alpha_\nu(\bt^{\nu}_{\so})
\gamma_{\nu}(\bt^{\nu}_{\st},\bt^{\nu}_{\so})}{f_{[\nu+1]}(\bt^{\nu+1},\bt^{\nu}_{\so})}.
\ee

Note that  in \eqref{T1jactBV} the
operators $T_{1,j}(z)$ act onto $\mathbb{B}(\bt)$, while in \eqref{recBV-G01} these operators act onto
$\mathbb{B}(\bigr\{\bt^1\bigr\};\bigl\{\bt^k_{\st}\bigr\}_{2}^{j-1};\bigl\{\bt^k\bigr\}_{j}^{N})$. Therefore, we can directly
use the action formula \eqref{T1jactBV} for $j=2$ only.
 For $j>2$  we should replace in \eqref{alj}
and \eqref{Hqj} the sets $\bt^2,\dots,\bt^{j-1}$ with the subsets $\bt^2_{\st},\dots,\bt^{j-1}_{\st}$ before
substituting \eqref{T1jactBV} into recursion \eqref{recBV-G01}.

We look for the terms in the formulas
\eqref{alj} and \eqref{Hqj} where we should do the replacement $\{\bt^2,\dots,\bt^{j-1}\}\to\{\bt^2_{\st},\dots,\bt^{j-1}_{\st}\}$.
The sets  $\{\bt^2,\dots,\bt^{j-1}\}$ appear  only in the factors
$h(\bt^m,z)^{[j]}$ and $h(\bt^m_{\st},z)^{[q]-[j]}$, and provided that $m\in\{2,\dots,j-1\}$. This implies that for $m=1$ there is no replacement to do. For $m>1$, we have $[j]=1$, because $j>m$, and $[q]=[j]$, because $q>j$.
Then, the factor $h(\bt^m_{\st},z)^{[q]-[j]}$ drops out, and we should only replace
$h(\bt^m,z)^{[j]}\to h(\bt_{\st}^m,z)^{[j]}$.



Thus, we arrive at the following action formula:
\begin{multline}\label{tT1jactBV}
T_{1,j}(z)\mathbb{B}(\bigr\{\bt^1\bigr\};\bigl\{\bt^k_{\st}\bigr\}_{2}^{j-1};\bigl\{\bt^k\bigr\}_{j}^{N})
= \tilde\eta_j\mathbb{B}(\bigr\{z,\bt^1\bigr\};\bigl\{z,\bt^k_{\st}\bigr\}_{2}^{j-1};\bigl\{\bt^k\bigr\}_{j}^{N})\\
+\sum_{q=j+1}^{N+1}\sum_{\text{part}(\bt^j,\dots,\bt^{q-1})}\tilde H_{q,j}(\text{part})
\mathbb{B}(\bigr\{z,\bt^1\bigr\};\{z,\bt^k_{\st}\}_2^{q-1};\{\bt^k\}_q^{N}),
\end{multline}
where
\be{talj}
\tilde \eta_j=\lambda_j(z)f_{[j]}(\bt^j,z)h(\bt^m_{\st},z)^{[j]}h(\bt^m_{\so},z)^{\delta_{m,1}},
\ee
and
\begin{equation}\label{tHqj}
\tilde H_{q,j}(\text{part})=f_{[q]}(\bt^q,z)h(\bt^m_{\st},z)^{[q]}h(\bt^m_{\so},z)^{\delta_{m,1}} \lambda_q(z)
g_{[j]}(z,\bt_{\so}^{q-1})\prod_{\nu=j+1}^{q-1}g_{[\nu]}(\bt^{\nu}_{\so},\bt^{\nu-1}_{\so})
\prod_{\nu=j}^{q-1}\Omega_\nu.
\end{equation}

Now everything is ready for substituting the action formula \eqref{tT1jactBV} into recursion \eqref{recBV-G01}. Let
\begin{equation}\label{defTB}
\mathbb{X}=
\sum_{j=2}^{N+1}T_{1,j}(z)\sum_{\text{\rm part}(\bt^2,\dots,\bt^{j-1})}
\frac{\prod_{\nu=2}^{j-1}g_{[\nu]}(\bt^{\nu}_{\so},\bt^{\nu-1}_{\so})\Omega_\nu}
{\lambda_2(z)h(\bt^1,z)^{\delta_{m,1}}f_{[2]}(\bt^{2},z)}
\mathbb{B}(\bigr\{\bt^1\bigr\};\bigl\{\bt^k_{\st}\bigr\}_{2}^{j-1};\bigl\{\bt^k\bigr\}_{j}^{N}) .
\end{equation}
It is easy to see that $\mathbb{X}$ is nothing else but the r.h.s. of recursion \eqref{recBV-G01}. Thus, our goal is to show that
$\mathbb{X}=\mathbb{B}(\bigr\{z,\bt^1\bigr\};\bigl\{\bt^k\bigr\}_{2}^{N})$.
Substituting  \eqref{tT1jactBV} into \eqref{defTB} we obtain
\begin{multline}\label{BV-rec}
\mathbb{X}=
\sum_{j=2}^{N+1}\sum_{\text{part}(\bt^2,\dots,\bt^{j-1})}
\frac{\tilde\eta_j \prod_{\nu=2}^{j-1}g_{[\nu]}(\bt^{\nu}_{\so},\bt^{\nu-1}_{\so})\Omega_\nu}{\lambda_2(z)h(\bt^1,z)^{\delta_{m,1}}f_{[2]}(\bt^{2},z)}
\mathbb{B}(\bigr\{z,\bt^1\bigr\};\bigl\{z,\bt^k_{\st}\bigr\}_{2}^{j-1};\bigl\{\bt^k\bigr\}_{j}^{N})\\
+\sum_{j=2}^{N+1}\sum_{q=j+1}^{N+1}\sum_{\text{part}(\bt^2,\dots,\bt^{q-1})}
\frac{\tilde H_{q,j}(\text{part}) \prod_{\nu=2}^{j-1}g_{[\nu]}(\bt^{\nu}_{\so},\bt^{\nu-1}_{\so})\Omega_\nu}{\lambda_2(z)h(\bt^1,z)^{\delta_{m,1}}f_{[2]}(\bt^{2},z)}
\mathbb{B}(\bigr\{z,\bt^1\bigr\};\{z,\bt^k_{\st}\}_2^{q-1};\{\bt^k\}_q^{N}).
\end{multline}
It is convenient to divide $\mathbb{X}$ into three contributions
\be{B-B3}
\mathbb{X}=\mathbb{X}^{(1)}+\mathbb{X}^{(2)}+\mathbb{X}^{(3)}.
\ee
The first term $\mathbb{X}^{(1)}$ corresponds to $j=2$ in the first line of \eqref{BV-rec}:
\be{B1}
\mathbb{X}^{(1)}=\frac{\tilde\eta_2 \mathbb{B}(\bigr\{z,\bt^1\bigr\};\bigl\{\bt^k\bigr\}_{2}^{N})}{\lambda_2(z)h(\bt^1,z)^{\delta_{m,1}}f_{[2]}(\bt^{2},z)}.
\ee
Substituting here $\tilde\eta_2$ we see that
\be{Opla}
\mathbb{X}^{(1)}=\mathbb{B}(\bigr\{z,\bt^1\bigr\};\bigl\{\bt^k\bigr\}_{2}^{N}).
\ee

The contribution $\mathbb{X}^{(2)}$ includes the terms with $j>2$ from the first line of \eqref{BV-rec}. The contribution $\mathbb{X}^{(3)}$ comes from the second line of \eqref{BV-rec}. Consider $\mathbb{X}^{(3)}$ changing the order of
summation and substituting there \eqref{tHqj}. We have
\begin{multline}\label{B3}
\mathbb{X}^{(3)}=
\sum_{q=3}^{N+1}\sum_{j=2}^{q-1}\sum_{\text{part}(\bt^2,\dots,\bt^{q-1})}
\frac{\lambda_q(z)f_{[q]}(\bt^q,z)h(\bt^m_{\st},z)^{[q]} h(\bt^m_{\so},z)^{\delta_{m,1}} }
{\lambda_2(z)h(\bt^1,z)^{\delta_{m,1}}f_{[2]}(\bt^{2},z)}\\
\times\frac{g(z,\bt_{\so}^{q-1})}{g(\bt^{j}_{\so},\bt^{j-1}_{\so})}\left(\prod_{\nu=2}^{q-1}g_{[\nu]}(\bt^{\nu}_{\so},\bt^{\nu-1}_{\so})\Omega_\nu\right)\;
\mathbb{B}(\bigr\{z,\bt^1\bigr\};\{z,\bt^k_{\st}\}_2^{q-1};\{\bt^k\}_q^{N}).
\end{multline}
The sum over $j$ can be easily computed
\be{chain}
\sum_{j=2}^{q-1}\frac{1}{g(\bt^{j}_{\so},\bt^{j-1}_{\so})}=\frac1c\sum_{j=2}^{q-1}(\bt^{j}_{\so}-\bt^{j-1}_{\so})=
\frac1c(\bt^{q-1}_{\so}-\bt^{1}_{\so})=-1/g(z,\bt^{q-1}_{\so}),
\ee
and we recall that by definition $\bt^{1}_{\so}=z$.
Thus,
\begin{multline}\label{B3-2}
\mathbb{X}^{(3)}=
-\sum_{q=3}^{N+1}\sum_{\text{part}(\bt^2,\dots,\bt^{q-1})}
\frac{\lambda_q(z)f_{[q]}(\bt^q,z)h(\bt^m_{\st},z)^{[q]}  }
{\lambda_2(z)h(\bt^1_{\st},z)^{\delta_{m,1}}f_{[2]}(\bt^{2},z)}
\prod_{\nu=2}^{q-1}g_{[\nu]}(\bt^{\nu}_{\so},\bt^{\nu-1}_{\so})\Omega_\nu\\
\times
\mathbb{B}(\bigr\{z,\bt^1\bigr\};\{z,\bt^k_{\st}\}_2^{q-1};\{\bt^k\}_q^{N}).
\end{multline}

On the other hand, the contribution $\mathbb{X}^{(2)}$ is
\begin{multline}\label{B2}
\mathbb{X}^{(2)}=
\sum_{j=3}^{N+1}\sum_{\text{part}(\bt^2,\dots,\bt^{j-1})}
\frac{\lambda_j(z)f_{[j]}(\bt^j,z)h(\bt^m_{\st},z)^{[j]}  }
{\lambda_2(z)h(\bt^1_{\st},z)^{\delta_{m,1}}f_{[2]}(\bt^{2},z)}
\prod_{\nu=2}^{j-1}g_{[\nu]}(\bt^{\nu}_{\so},\bt^{\nu-1}_{\so})\Omega_\nu\\
\times
\mathbb{B}(\bigr\{z,\bt^1\bigr\};\{z,\bt^k_{\st}\}_2^{j-1};\{\bt^k\}_j^{N}).
\end{multline}
Comparing \eqref{B2} and \eqref{B3-2} we see that they cancel each other. Thus, $\mathbb{X}=\mathbb{B}(\bigr\{z,\bt^1\bigr\};\bigl\{\bt^k\bigr\}_{2}^{N})$. \qed

\subsection{Proofs of proposition~\ref{C2-recBV}\label{SS-PoC}}

Let us derive now recursion \eqref{recBV-G02pr} starting with  \eqref{recBV-G01} and using morphism \eqref{morphism}. Since the
mapping \eqref{morphism} relates two different Yangians $Y(\mathfrak{gl}(m|n))$ and $Y(\mathfrak{gl}(n|m))$, we use here additional
superscripts for the functions $g(u,v)$, $f(u,v)$, $\gamma(u,v)$, and $\hg(u,v)$. For example, notation $f^{m|n}_{[\nu]}(u,v)$
means that the function $f_{[\nu]}(u,v)$ is defined with respect to $Y(\mathfrak{gl}(m|n))$:
\be{fmn}
f^{m|n}_{[\nu]}(u,v)=\left[\begin{array}{l} f(u,v),\qquad \nu\le m,\\ f(v,u),\qquad \nu > m.
\end{array}\right.
\ee
At the same time the notation $f^{n|m}_{[\nu]}(u,v)$
means that the function $f_{[\nu]}(u,v)$ is defined with respect to $Y(\mathfrak{gl}(n|m))$:
\be{fnm}
f^{n|m}_{[\nu]}(u,v)=\left[\begin{array}{l} f(u,v),\qquad \nu\le n,\\ f(v,u),\qquad \nu > n.
\end{array}\right.
\ee
The other rational functions should be understood similarly. It is easy to see that
\begin{equation}\label{iden}
\begin{aligned}
&  g^{m|n}_{[\nu]}(u,v) = g^{n|m}_{[N + 2 - \nu]}(v,u),\\
&  f^{m|n}_{[\nu]}(u,v) = f^{n|m}_{[N + 2 - \nu]}(v,u),\\
& \gamma^{m|n}_{\nu}(u,v) = \hg^{n|m}_{N + 1 - \nu}(v,u).
\end{aligned}
\end{equation}

Let us act with $\varphi$ onto \eqref{recBV-G01}. Due to \eqref{morphism}--\eqref{morph-BV} we have
\be{mor-Tl}
\varphi\left( \frac{T^{m|n}_{1,j}(z)}{\lambda_2(z)}\right)=(-1)^{[j]}\frac{T^{n|m}_{N+2-j,N+1}(z)}{\lambda_N(z)},
\ee
\be{mor-BV1}
\varphi\left( \mathbb{B}^{m|n}(\bigr\{z,\bt^1\bigr\};\bigl\{\bt^k\bigr\}_{2}^{N})\right)=(-1)^{r_m+\delta_{m,1}}
\frac{\mathbb{B}^{n|m}(\bigl\{\bt^k\bigr\}_{N}^{2};\bigr\{z,\bt^1\bigr\})}
{\alpha_N(z)\prod_{k=1}^N\alpha_{N+1-k}(\bt^k)},
\ee
and
\be{mor-BV2}
\varphi\left( \mathbb{B}^{m|n}(\bigr\{\bt^1\bigr\};\bigl\{\bt^k_{\st}\bigr\}_{2}^{j-1};\bigl\{\bt^k\bigr\}_{j}^{N})
\prod_{\nu=2}^{j-1}\alpha_\nu(\bt^\nu_{\so})\right)
=(-1)^{r_m+\delta_{m,1}+[j]}\frac{\mathbb{B}^{n|m}(\bigr\{\bt^k\bigr\}_N^{j};\bigl\{\bt^k_{\st}\bigr\}_{j-1}^{2};\bt^{1})}
{\prod_{k=1}^N\alpha_{N+1-k}(\bt^k)}
\ee
Thus, the action of the morphism $\varphi$ onto \eqref{recBV-G01} gives
\begin{multline}\label{phi-rec1}
\mathbb{B}^{n|m}(\bigl\{\bt^k\bigr\}_{N}^{2};\bigr\{z,\bt^1\bigr\})
=\sum_{j=2}^{N+1}\frac{T_{N+2-j,N+1}(z)}{\lambda_{N+1}(z)}
\sum_{\text{\rm part}(\bt^2,\dots,\bt^{j-1})}
\mathbb{B}^{n|m}(\bigr\{\bt^k\bigr\}_N^{j};\bigl\{\bt^k_{\st}\bigr\}_{j-1}^{2};\bt^{1})
\\
\times
\frac{\prod_{\nu=2}^{j-1}g^{m|n}_{[\nu]}(\bt^{\nu}_{\so},\bt^{\nu-1}_{\so})\gamma^{m|n}_{\nu}(\bt^{\nu}_{\st},\bt^{\nu}_{\so})}
     {h(\bt^1,z)^{\delta_{m,1}}\prod_{\nu=1}^{j-1}f^{m|n}_{[\nu+1]}(\bt^{\nu+1},\bt^{\nu}_{\so})}.
\end{multline}
Using the relations \eqref{iden}
and the trivial identity $\delta_{m,1} = \delta_{n,N}$
we recast \eqref{phi-rec1} as
\begin{multline}\label{phi-rec2}
\mathbb{B}^{n|m}(\bigl\{\bt^k\bigr\}_{N}^{2};\bigr\{z,\bt^1\bigr\})
=\sum_{j=2}^{N+1}\frac{T_{N+2-j,N+1}(z)}{\lambda_{N+1}(z)}
\sum_{\text{\rm part}(\bt^2,\dots,\bt^{j-1})}
\mathbb{B}^{n|m}(\bigr\{\bt^k\bigr\}_N^{j};\bigl\{\bt^k_{\st}\bigr\}_{j-1}^{2};\bt^{1})
\\
\times
\frac{\prod_{\nu=2}^{j-1}g^{n|m}_{[N+2 - \nu]}(\bt^{\nu-1}_{\so},\bt^{\nu}_{\so})\hat\gamma^{n|m}_{N+1-\nu}(\bt^{\nu}_{\so},\bt^{\nu}_{\st})}
     {h(\bt^1,z)^{\delta_{n,N}}\prod_{\nu=1}^{j-1}f^{n|m}_{[N+1-\nu]}(\bt^{\nu}_{\so},\bt^{\nu+1})}.
\end{multline}
Finally, relabeling the sets of the Bethe parameters $\bar t^k \to \bar t^{N+1-k}$  and changing $\nu\to N+1-\nu$ we obtain
\begin{multline}\label{phi-rec3}
\mathbb{B}^{n|m}(\bigl\{\bt^k\bigr\}_{1}^{N-1};\{ z, \bar t^N \})
=\sum_{j=1}^{N}\frac{T_{j,N+1}(z)}{\lambda_{N+1}(z)}
\sum_{\text{\rm part}(\bt^j,\dots,\bt^{N-1})}
\mathbb{B}^{n|m}(\bigr\{\bt^k\bigr\}_1^{j-1};\bigl\{\bt^k_{\st}\bigr\}_{j}^{N-1};\bt^{N})
\\
\times
\frac{\prod_{\nu=j}^{N-1}g^{n|m}_{[\nu+1]}(\bt^{\nu+1}_{\so},\bt^{\nu}_{\so})\hat\gamma^{n|m}_{\nu}(\bt^{\nu}_{\so},\bt^{\nu}_{\st})}
     {h(\bt^N,z)^{\delta_{n,N}}\prod_{\nu=j}^{N}f^{n|m}_{[\nu]}(\bt^{\nu}_{\so},\bt^{\nu-1})}.
\end{multline}
It remains to replace $m\leftrightarrow n$, and we arrive at \eqref{recBV-G02pr}.\qed

\subsection{Proof of recursion for dual Bethe vectors\label{SS-RfdBV}}

To obtain recursion for dual Bethe vectors it is enough to act with antimorphism \eqref{antimo} onto recursions
\eqref{recBV-G01} and \eqref{recBV-G02pr}. Consider in details the action of $\Psi$ onto \eqref{recBV-G01}.

Acting with $\Psi$ on the lhs of \eqref{recBV-G01} we obtain a dual vector $\mathbb{C}(\bigr\{z,\bt^1\bigr\};\bigl\{\bt^k\bigr\}_{2}^{N})$
due to \eqref{antimoBV}. In the rhs we have
\be{ActPsiTBV}
\Psi(T_{1,j}\mathbb{B})=(-1)^{[j][\mathbb{B}]}\mathbb{C}\;T_{j,1}.
\ee
The parity of the Bethe vector can be determined via the coloring arguments. Recall that Bethe vectors are polynomials in the operators $T_{i,j}$
acting on the vector $|0\rangle$, and all the terms of these polynomials have the same coloring.
Due to the general rule, a quasiparticle
of the color $m$ can be created by the operators $T_{i,j}$ with $i\le m$ and $j>m$. Hence, all these operators
are odd, because $[i]=0$ for $i\le m$ and $[j]=1$ for $j>m$. On the other hand, the action of an even operator $T_{i,j}$ cannot create a quasiparticle
of the color $m$ due to similar arguments.   Thus, if a Bethe vector has a coloring $\{r_1,\dots,r_N\}$, then all the terms
of the polynomial in $T_{i,j}$ contain exactly $r_m$ odd operators, where $r_m=\#\bt^m$. Thus,
$\bigl[\mathbb{B}(\bt)\bigr]=r_m,\,\mod 2$.

In the case under consideration we should find  the number $r'_m$ of the odd operators
in the Bethe vector $\mathbb{B}(\bigr\{\bt^1\bigr\};\bigl\{\bt^k_{\st}\bigr\}_{2}^{j-1};\bigl\{\bt^k\bigr\}_{j}^{N})$. Let $r_m=\#\bt^m$ in the
original vector $\mathbb{B}(\bt)$. If $m=1$, then $r'_m=r_m$. If $1<m<j$, then $r'_m=r_m-1$. Finally, if $m\ge j$, then $r'_m=r_m$.
All these cases can be described by the formula $r'_m=r_m-[j]+\delta_{m,1}$. Thus, we obtain
\begin{multline}\label{recdBV-antimo}
\mathbb{C}(\bigr\{z,\bt^1\bigr\};\bigl\{\bt^k\bigr\}_{2}^{N})=
\sum_{j=2}^{N+1}
\sum_{\text{\rm part}(\bt^2,\dots,\bt^{j-1})}\mathbb{C}(\bigr\{\bt^1\bigr\};\bigl\{\bt^k_{\st}\bigr\}_{2}^{j-1};\bigl\{\bt^k\bigr\}_{j}^{N})
\frac{T_{j,1}(z)}{\lambda_2(z)} (-1)^{[j]r'_m} \\
\times
\frac{\prod_{\nu=2}^{j-1}\alpha_\nu(\bt^\nu_{\so})g_{[\nu]}(\bt^{\nu}_{\so},\bt^{\nu-1}_{\so})
\gamma_{\nu}(\bt^{\nu}_{\st},\bt^{\nu}_{\so})}{h(\bt^1,z)^{\delta_{m,1}}\prod_{\nu=1}^{j-1}f_{[\nu+1]}(\bt^{\nu+1},\bt^{\nu}_{\so})},
\end{multline}
where $r'_m=r_m-[j]+\delta_{m,1}$.

This expression can be slightly simplified. Recall that $\hg_{i}(x,y)=(-1)^{\delta_{m,i}}\gamma_i(x,y)$. Thus, changing
$\gamma_\nu(\bt^{\nu}_{\st},\bt^{\nu}_{\so})\to \hg_\nu(\bt^{\nu}_{\st},\bt^{\nu}_{\so})$ in \eqref{recdBV-antimo} we obtain
\be{prodgam}
\prod_{\nu=2}^{j-1}\gamma_\nu(\bt^{\nu}_{\st},\bt^{\nu}_{\so})
=(-1)^{([j]-[2])r'_m}\prod_{\nu=2}^{j-1} \hg_\nu(\bt^{\nu}_{\st},\bt^{\nu}_{\so}).
\ee
It remains to observe that $[2]=\delta_{m,1}$. Thus, substituting \eqref{prodgam} into \eqref{recdBV-antimo} and replacing
the sets $\bt^k$ with $\bs^k$ we arrive at \eqref{recdBV-G01}. Recursion \eqref{C-recDBV2} can be obtained exactly in the same way.

\section{Proof of the sum formula for the scalar product\label{S-SF}}

\subsection{How the scalar product depends on the vacuum eigenvalues $\lambda_i(z)$\label{SS-HSCPD}}

In this section, we investigate the functional dependence of the scalar product on the functions $\alpha_i$.
 Proposition~\ref{P-al-dep} states that the Bethe parameters from the sets $\bs^i$ and $\bt^i$ can be the arguments of the functions
$\alpha_i$ only. In other words, the scalar product does not depend on $\alpha_i(s^\ell_k)$ or $\alpha_i(t^\ell_k)$ with $\ell\ne i$.
%
%

We prove this statement via induction over $N=m+n-1$. For $N=1$ it becomes obvious. Assume that it is valid for some $N-1$
 and consider the scalar product of the vectors $\mathbb{C}^{m|n}(\bs)$ and $\mathbb{B}^{m|n}(\bt)$ with $m+n-1=N$.
Observe that we added superscripts to the Bethe
vectors in order to distinguish them from the vectors corresponding to $\mathfrak{gl}(m-1|n)$ algebra.
We first prove that the scalar product  does not depend
on the functions $\alpha_i(s^\ell_k)$ with $\ell\ne i$ for $i=2,\dots,N$.

Successive application of the recursion \eqref{recdBV-G01} allows one to express a dual Bethe vector $\mathbb{C}^{m|n}(\bs)$
in terms of dual Bethe vectors $\mathbb{C}^{m-1|n}(\bar\sigma)$. Schematically this expression can be written in the following form
\begin{equation}\label{recdBV-Big}
\mathbb{C}^{m|n}(\bs)=
\sum_{j_1,\dots,j_{r_1}=2}^{m+n}
\sum_{\{\bar\sigma^2,\dots,\bar\sigma^{N}\}}\Theta_{j_1,\dots,j_{r_1}}^{(\bs)}(\bar\sigma)
\mathbb{C}^{m-1|n}(\bigl\{\bar\sigma\bigr\}_{2}^{N})
\frac{T_{j_1,1}(s^1_1)\dots T_{j_{r_1},1}(s^{1}_{r_1})}{\lambda_2(\bs^1)}.
\end{equation}
Here $r_1=\#\bs^1$ and $\bar\sigma^i\subset\bs^i$ for $i=2,\dots,N$. The sum is taken over multi-index $\{j_1,\dots,j_{r_1}\}$.
Every term of this sum contains also a sum over partitions
of the sets $\bs^2,\dots,\bs^N$ into subsets $\bar\sigma^2,\dots,\bar\sigma^N$ and their complementary subsets. The factors $\Theta_{j_1,\dots,j_{r_1}}^{(\bs)}(\bar\sigma)$ are some numerical coefficients whose explicit form is
not essential. It is important, however, to note that in
\eqref{recdBV-G01} they depend on $\alpha_i(s^i_k)$ with $i=2,\dots,N$ and do not depend on the functions $\alpha_i$ with other arguments.
%
%

Let us multiply \eqref{recdBV-Big} from the right by a Bethe vector $\mathbb{B}^{m|n}(\bt)$ and act with the operators
$T_{j_{p},1}(s^{1}_{p})$ onto this vector. Due to the results of \cite{HutLPRS17a} the action of any operator $T_{ij}(z)$
onto the Bethe vector $\mathbb{B}^{m|n}(\bt)$ gives a linear combination of new Bethe vectors $\mathbb{B}^{m|n}(\bar\tau)$,
such that $\bar\tau=\{\bar\tau^1,\dots,\bar\tau^N\}$ and $\bar\tau^i\subset \{\bt^i\cup z\}$. In the case under consideration
each of the operators $T_{j_{p},1}(s^{1}_{p})$ annihilates a particle of  color $1$. Hence, the total action
of $T_{j_1,1}(s^1_1)\dots T_{j_{r_1},1}(s^{1}_{r_1})$ annihilates all the particles of  color $1$ in the vector $\mathbb{B}^{m|n}(\bt)$.
Thus, after this action the Bethe vector $\mathbb{B}^{m|n}(\bt)$ turns into $\mathbb{B}^{m-1|n}(\bar\tau)$, where
$\bar\tau=\{\bar\tau^2,\dots,\bar\tau^N\}$ and $\bar\tau^i\subset \{\bt^i\cup\bs^1\}$
\begin{equation}\label{actTBV-Big}
\frac{T_{j_1,1}(s^1_1)\dots T_{j_{r_1},1}(s^{1}_{r_1})}{\lambda_2(\bs^1)}\mathbb{B}^{m|n}(\bt)=
\sum_{\{\bar\tau^2,\dots,\bar\tau^{N}\}}\Theta^{(\bt)}(\bar\tau)
\mathbb{B}^{m-1|n}(\bigl\{\bar\tau^k\bigr\}_{2}^{N}).
\end{equation}
Here the coefficients $\Theta^{(\bt)}(\bar\tau)$ of the linear combination depend on the original sets $\bt^k$ and subsets $\bar\tau^k$.
They involve the functions $\alpha_i$ whose arguments belong to the set $\{\bs^1\cup\bt\}$. Therefore, the factors $\Theta^{(\bt)}(\bar\tau)$
do not depend on $\alpha_j(s^i_k)$ with $i,j=2,\dots,N$.
%
%

Thus, we obtain a recursion for the scalar product
\begin{equation}\label{recSP-Big}
\mathbb{C}^{m|n}(\bs)\mathbb{B}^{m|n}(\bt)=
\sum_{\substack{\{\bar\sigma^2,\dots,\bar\sigma^{N}\} \\\{\bar\tau^2,\dots,\bar\tau^{N}\}  }}
\Theta_{j_1,\dots,j_{r_1}}^{(\bs)}(\bar\sigma)\Theta^{(\bt)}(\bar\tau)\;
\mathbb{C}^{m-1|n}(\bigl\{\bar\sigma^k\bigr\}_{2}^{N})\mathbb{B}^{m-1|n}(\bigl\{\bar\tau^k\bigr\}_{2}^{N}),
\end{equation}
where $\bar\sigma^k\subset\bs^k$ and $\bar\tau^k\subset\{\bs^1\cup\bt^k\}$. The sum is taken over subsets $\bar\sigma^k$ and $\bar\tau^k$.

Due to the induction assumption, the scalar product $\mathbb{C}^{m-1|n}(\bigl\{\bar\sigma^k\bigr\}_{2}^{N})\mathbb{B}^{m-1|n}(\bigl\{\bar\tau^k\bigr\}_{2}^{N})$
depends on the functions $\alpha_i$ with arguments $\sigma^i_k$ and $\tau^i_k$. Since $\sigma^i_k\in\bs^i$, we conclude that the Bethe parameters
$s^i_k$ for $i=2,\dots,N$ can become the arguments of the functions $\alpha_i$ only. The numerical coefficients $\Theta_{j_1,\dots,j_{r_1}}^{(\bs)}(\bar\sigma)$ and $\Theta^{(\bt)}(\bar\tau)$
do not break this type of dependence.
Thus, we prove that in the scalar product $\mathbb{C}^{m|n}(\bs)\mathbb{B}^{m|n}(\bt)$ the  Bethe parameters
$s^i_k$ with $i=2,\dots,N$ can become the arguments of the functions $\alpha_i$ only.

Due to the symmetry \eqref{SP-def1}, an  analogous property holds for the Bethe parameters $\bt^i$ with $i=2,\dots,N$. Namely,
these parameters can be the arguments of the functions $\alpha_i$ only.

It remains to prove that the Bethe parameters from the sets $\bs^1$ and $\bt^1$ can be the arguments of the function $\alpha_1$. For this we use
the second recursion for the dual Bethe vector \eqref{C-recDBV2} and repeat all the considerations above. Then we find that the Bethe parameters
$s^i_k$ with $i=1,\dots,N-1$ can become the arguments of the functions $\alpha_i$ only. Then, the use of \eqref{SP-def1} completes the proof of
proposition~\ref{P-al-dep}.\qed

\subsection{Proof of the sum formula\label{S-RF}}

Consider a composite model, in which the monodromy matrix $T(u)$ is presented as a product of two partial monodromy
matrices \cite{IzeK84,HutLPRS17a,Fuk17,FukS17}:
\be{T-TT}
T(u)=T^{(2)}(u)T^{(1)}(u).
\ee
Within the framework of the composite model, it is assumed that the matrix elements of every $T^{(l)}(u)$ ($l=1,2$) act in some Hilbert space
$\mathcal{H}^{(l)}$,  such that $\mathcal{H}=\mathcal{H}^{(1)}\otimes \mathcal{H}^{(2)}$. Each of $T^{(l)}(u)$
satisfies the $RTT$-relation \eqref{RTT} and has its own pseudovacuum vector $|0\rangle^{(l)}$ and
dual vector $\langle0|^{(l)}$, such that $|0\rangle= |0\rangle^{(1)}\otimes|0\rangle^{(2)}$ and  $\langle0|=\langle0|^{(1)}\otimes \langle0|^{(2)}$.
Since the operators $T_{i,j}^{(2)}(u)$ and $ T_{k,l}^{(1)}(v)$ act in different spaces, they supercommute with each other. We assume that
\be{eigen}
\begin{aligned}
T_{i,i}^{(l)}(u)|0\rangle^{(l)}&= \lambda_{i}^{(l)}(u)|0\rangle^{(l)}, \\
\langle0|^{(l)}T_{i,i}^{(l)}(u)&= \lambda_{i}^{(l)}(u)\langle0|^{(l)},
\end{aligned}
\qquad i=1,\dots,m+n,\qquad l=1,2,
\ee
where $\lambda_{i}^{(l)}(u)$ are new free functional parameters.
 We also introduce
\be{rk}
\alpha_{k}^{(l)}(u)=\frac{\lambda_{k}^{(l)}(u)}{\lambda_{k+1}^{(l)}(u)}, \qquad l=1,2, \qquad k=1,\dots,N.
\ee
Obviously
\be{lr}
\lambda_{i}(u)=\lambda_{i}^{(1)}(u)\lambda_{i}^{(2)}(u), \qquad
\alpha_{k}(u)=\alpha_{k}^{(1)}(u)\alpha_{k}^{(2)}(u).
\ee

The partial monodromy matrices $T^{(l)}(u)$ have the corresponding Bethe vectors $\mathbb{B}^{(l)}(\bt)$ and
dual Bethe vectors $\mathbb{C}^{(l)}(\bs)$. A Bethe vector of the total monodromy matrix $T(u)$ can be expressed
in terms partial Bethe vectors $\mathbb{B}^{(l)}(\bt)$ via {\it coproduct formula}\footnote{%
The terminology {\it coproduct formula} is used for historical reason, because \eqref{BB11} was derived for the first time in
\cite{HutLPRS17a} (see also \cite{KhP-Kyoto} for the non-graded case) as a property of the Bethe vectors induced by  the Yangian coproduct. }
\cite{Fuk17,HutLPRS17a}
\begin{equation}\label{BB11}
 \mathbb{B}(\bar t)=\sum\frac{
 \prod_ {\nu=1}^{N} \alpha^{(2)}_{\nu}(\bt^\nu_{\rm i}) \gamma_{\nu}(\bt^\nu_{\rm ii},\bt^\nu_{\rm i})}
   { \prod_{\nu=1}^{N-1} f_{[\nu+1]}(\bt^{\nu+1}_{\rm ii},\bt^\nu_{\rm i})}\;
 \mathbb{B}^{(1)}(\bt_{\rm i})  \otimes
 \mathbb{B}^{(2)}(\bt_{\rm ii}) .
\end{equation}
Here all the sets of the Bethe parameters $\bt^\nu$  are divided into two subsets $\bt^\nu\Rightarrow\{\bt^\nu_{\rm i}, \bt^\nu_{\rm ii} \}$,
and the sum is taken over all possible partitions.

Similar formula exists for the dual Bethe vectors $\mathbb{C}(\bs)$ (see appendix~\ref{SS-CPFSP})
\begin{equation}\label{BC1}
 \mathbb{C}(\bar s)=\sum\frac{
 \prod_ {\nu=1}^{N}  \alpha^{(1)}_{\nu}(\bs^\nu_{\rm ii})\gamma_{\nu}(\bs^\nu_{\rm i},\bs^\nu_{\rm ii})}
{\prod_{\nu=1}^{N-1} f_{[\nu+1]}(\bs^{\nu+1}_{\rm i},\bs^\nu_{\rm ii})   }\;
 \mathbb{C}^{(2)}(\bs_{\rm ii}) \otimes \mathbb{C}^{(1)}(\bs_{\rm i}),
\end{equation}
where the sum is organized in the same way as in \eqref{BB11}.

Then the scalar product of the total Bethe vectors $ \mathbb{C}(\bs)$ and $\mathbb{B}(\bt)$ takes the form
\begin{equation}\label{Sab}
S(\bs|\bt)=\sum \frac{\prod_ {\nu=1}^{N}  \alpha^{(1)}_{\nu}(\bs^\nu_{\rm ii})\alpha^{(2)}_{\nu}(\bt^\nu_{\rm i})
\gamma_{\nu}(\bs^\nu_{\rm i},\bs^\nu_{\rm ii})  \gamma_{\nu}(\bt^\nu_{\rm ii},\bt^\nu_{\rm i})}
{\prod_{\nu=1}^{N-1} f_{[\nu+1]}(\bs^{\nu+1}_{\rm i},\bs^\nu_{\rm ii}) f_{[\nu+1]}(\bt^{\nu+1}_{\rm ii},\bt^\nu_{\rm i})   }\;
S^{(1)}(\bs_{\rm i}|\bt_{\rm i})S^{(2)}(\bs_{\rm ii}|\bt_{\rm ii}),
\end{equation}
where
\be{S-CBl}
S^{(1)}(\bs_{\rm i}|\bt_{\rm i})=\mathbb{C}^{(1)}(\bs_{\rm i})\mathbb{B}^{(1)}(\bt_{\rm i}),\qquad
S^{(2)}(\bs_{\rm ii}|\bt_{\rm ii})=\mathbb{C}^{(2)}(\bs_{\rm ii})\mathbb{B}^{(2)}(\bt_{\rm ii}).
\ee
Note that in this formula $\#\bs^\nu_{\rm i}=\#\bt^\nu_{\rm i}$, (and hence, $\#\bs^\nu_{\rm ii}=\#\bt^\nu_{\rm ii}$), otherwise the
scalar products $S^{(1)}$ and $S^{(2)}$ vanish. Let
$\#\bs^\nu_{\rm i}=\#\bt^\nu_{\rm i}=k'_\nu$, where $k'_\nu=0,1,\dots,r_\nu$. Then $\#\bs^\nu_{\rm ii}=\#\bt^\nu_{\rm ii}=r_\nu-k'_\nu$.

Now let us turn to equation \eqref{al-dep}. Our goal is to express the rational coefficients $W^{m|n}_{\text{\rm part}}$ in terms of the HC.
For this we use the fact that $W^{m|n}_{\text{\rm part}}$ are model independent. Therefore, we can find them in some special model whose
monodromy matrix satisfies the $RTT$-relation.

Let us fix some partitions of the Bethe parameters in \eqref{al-dep}:
$\bs^{\nu}\Rightarrow\{\bs^{\nu}_{\so},\bs^{\nu}_{\st}\}$ and $\bt^{\nu}\Rightarrow\{\bt^{\nu}_{\so},\bt^{\nu}_{\st}\}$ such that $\#\bs^{\nu}_{\so}=\#\bt^{\nu}_{\so}=k_\nu$, where $k_\nu=0,1,\dots,r_\nu$. Hence, $\#\bs^{\nu}_{\st}=\#\bt^{\nu}_{\st}=r_\nu-k_\nu$.
Consider a concrete model, in which\footnote{This choice of the functions $\alpha_k$ is always possible, for example, within the
framework of inhomogeneous model with spins in higher dimensional representations, in which inhomogeneities coincide with some of the Bethe parameters.}
\be{ll13}
\begin{aligned}
&\alpha_\nu^{(1)}(z)=0,\quad\text{if}\quad z\in\bs^\nu_{\st};\\
&\alpha_\nu^{(2)}(z)=0, \quad\text{if}\quad z\in\bt^\nu_{\so}.
\end{aligned}
\ee
Due to \eqref{lr} these conditions imply
\be{al-zero}
\alpha_\nu(z)=0,\quad\text{if}\quad z\in\bs^\nu_{\st}\cup\bt^\nu_{\so}.
\ee
Then the scalar product is proportional to the coefficient $W^{m|n}_{\text{\rm part}}(\bs_{\so},\bs_{\st}|\bt_{\so},\bt_{\st})$, because
all other terms in the sum over partitions \eqref{al-dep} vanish due to the condition \eqref{al-zero}. Thus,
\begin{equation}\label{S-W}
S(\bs|\bt)= W^{m|n}_{\text{\rm part}}(\bs_{\so},\bs_{\st}|\bt_{\so},\bt_{\st})
  \prod_{k=1}^{N} \alpha_{k}(\bs^k_{\so}) \alpha_{k}(\bt^k_{\st}).
\end{equation}
On the other hand, \eqref{ll13} implies that a non-zero contribution in \eqref{Sab} occurs if and only if $\bs^\nu_{\rm ii}\subset\bs^\nu_{\so}$
and $\bt^\nu_{\rm i}\subset\bt^\nu_{\st}$. Hence, $r_\nu-k'_\nu\le k_\nu$ and $k'_\nu\le r_\nu-k_\nu$.
But this is possible if and only if $k'_\nu+k_\nu=r_\nu$. Thus, $\bs^\nu_{\rm ii}=\bs^\nu_{\so}$
and $\bt^\nu_{\rm i}=\bt^\nu_{\st}$. Then, for the complementary subsets we obtain $\bs^\nu_{\rm i}=\bs^\nu_{\st}$
and $\bt^\nu_{\rm ii}=\bt^\nu_{\so}$.
Thus, we arrive at
\begin{equation}\label{Sab-1}
S(\bs|\bt)=\frac{\prod_ {\nu=1}^{N}  \alpha^{(1)}_{\nu}(\bs^\nu_{\so})\alpha^{(2)}_{\nu}(\bt^\nu_{\st})
\gamma_{\nu}(\bs^\nu_{\st},\bs^\nu_{\so})  \gamma_{\nu}(\bt^\nu_{\so},\bt^\nu_{\st})}
{\prod_{\nu=1}^{N-1} f_{[\nu+1]}(\bs^{\nu+1}_{\st},\bs^\nu_{\so}) f_{[\nu+1]}(\bt^{\nu+1}_{\so},\bt^\nu_{\st})   }\;
 S^{(1)}(\bs_{\st}|\bt_{\st})S^{(2)}(\bs_{\so}|\bt_{\so}).
\end{equation}

It is easy to see that calculating the scalar product $S^{(1)}(\bs_{\st}|\bt_{\st})$ we should take only the term corresponding
to the conjugated HC. Indeed, all other terms are proportional to $\alpha_\nu^{(1)}(z)$ with $z\in\bs^\nu_{\st}$, therefore, they
vanish. Hence
\be{HC-22}
 S^{(1)}(\bs_{\st}|\bt_{\st})=\prod_{\nu=1}^N\alpha_\nu^{(1)}(\bt^\nu_{\st})\cdot
   \overline{Z}^{m|n}(\bs_{\st}|\bt_{\st}).
 \ee
Similarly, calculating the scalar product $S^{(2)}(\bs_{\so}|\bt_{\so})$ we should take only the term corresponding
to the  HC:
\be{HC-2}
S^{(2)}(\bs_{\so}|\bt_{\so})=
\prod_{\nu=1}^N\alpha_\nu^{(2)}(\bs^\nu_{\so})\;
\cdot Z^{m|n}(\bs_{\so}|\bt_{\so}).
\ee
Substituting this into \eqref{Sab-1} and using \eqref{lr}, \eqref{S-W} we arrive at
\begin{equation}\label{HypResh00}
W^{m|n}_{\text{\rm part}}(\bs_{\so},\bs_{\st}|\bt_{\so},\bt_{\st}) =  Z^{m|n}(\bs_{\so}|\bt_{\so}) \;\; \overline{Z}^{m|n}(\bs_{\st}|\bt_{\st})
 \frac{\prod_ {k=1}^{N}  \gamma_{k}(\bs^k_{\st},\bs^k_{\so}) \gamma_{k}(\bt^k_{\so},\bt^k_{\st})}
{\prod_{j=1}^{N-1} f_{[j+1]}(\bar s^{j+1}_{\st},\bar s^j_{\so})
 f_{[j+1]}(\bar t^{j+1}_{\so},\bar t^j_{\st})}.
\end{equation}
This expression obviously coincides with \eqref{HypResh} due to \eqref{Z-bZ}.

\section{Highest coefficient\label{S-HC}}

\subsection{Proof of the recursion for the Highest Coefficient\label{SS-PRHC}}

It follows from proposition~\ref{P-al-dep} that the scalar product is a sum, in which every term is proportional to
a product of the functions $\alpha_k$. Let us call a term {\it unwanted}, if  the corresponding product of the functions $\alpha_k$
contains at least one $\alpha_k( t^k_j)$, where $t^k_j\in\bt$. Respectively, a term is {\it wanted}, if all functions $\alpha_k$
depend on the Bethe parameters $s^k_j$ from the set $\bs$.

Below we consider some equations modulus
unwanted terms. In this case we use a symbol $\cong$. Thus, an equation of the type
$ lhs\cong rhs$ means that the $lhs$ is equal to the $rhs$ modulus unwanted terms.

Using the notion of unwanted terms one can redefine the HC  \eqref{HCdef} as follows:
\be{SP-HC}
S(\bs|\bt)\cong \prod_{k=1}^{N}\alpha_k(\bs^k)\cdot Z^{m|n}(\bs|\bt).
\ee
On the other hand, it follows from the explicit form of Bethe vectors \cite{HutLPRS17a} that
\be{BV-phi}
\mathbb{B}(\bt)\cong \widetilde{\mathbb{B}}(\bt)=\frac{ \bT_{1,2}(\bt^1)\dots \bT_{N,N+1}(\bt^N)|0\rangle}
{\prod_{j=1}^{N} \lambda_{j+1}(\bar t^j)
\prod_{j=1}^{N-1} f_{[j+1]}(\bar t^{j+1},\bar t^j)},
\ee
because all other terms in the Bethe vector contain factors $\alpha_{k}(t^k_j)$, and thus, they are unwanted. Hence, in order to
find the HC it is enough to consider a {\it reduced} scalar product $\tilde S(\bs|\bt)$
\be{tSP}
S(\bs|\bt)\cong \tilde S(\bs|\bt)=
\mathbb{C}(\bs)\widetilde{\mathbb{B}}(\bt).
\ee
%
%
%

In order to calculate the reduced scalar product \eqref{tSP} we can use the recursion \eqref{recdBV-G01}
for the dual Bethe vector $\mathbb{C}(\bs)$. We write it in the form
\begin{multline}\label{Rec-C1}
\mathbb{C}(\bs)=
\sum_{p=2}^{N+1}
\sum_{\text{\rm part}(\bs^2,\dots,\bs^{p-1})}\mathbb{C}(\bigl\{\bs^k_{\st}\bigr\}_{1}^{p-1};\bigl\{\bs^k\bigr\}_{p}^{N})
\frac{T_{p,1}(\bs^1_{\so})}{\lambda_2(\bs^1_{\so})} (-1)^{(r_1-1)\delta_{m,1}} \\
\times
\frac{\prod_{\nu=2}^{p-1}\alpha_\nu(\bs^\nu_{\so})g_{[\nu]}(\bs^{\nu}_{\so},\bs^{\nu-1}_{\so})
 \hg_{\nu}(\bs^{\nu}_{\st},\bs^{\nu}_{\so})}{h(\bs^1,\bs^1_{\so})^{\delta_{m,1}}\prod_{\nu=1}^{p-1}
 f_{[\nu+1]}(\bs^{\nu+1},\bs^{\nu}_{\so})}.
\end{multline}
Here the sum is taken over partitions of the sets $\bs^k\Rightarrow\{\bs^k_{\so},\bs^{k}_{\st}\}$ for $k=2,\dots,p$, such that
$\#\bs^k_{\so}=1$. The Bethe parameter $\bs^1_{\so}$ is fixed, and hence, the subset $\bs^1_{\st}$ also is fixed. There is no the sum over partitions of the
set $\bs^1$ in \eqref{Rec-C1}.

Thus, we obtain
\begin{multline}\label{Rec-tSP0}
\tilde S(\bs|\bt)=\sum_{p=2}^{N+1}
\sum_{\text{\rm part}(\bs^2,\dots,\bs^{p-1})} (-1)^{(r_1-1)\delta_{m,1}}
\mathbb{C}(\bigl\{\bs^k_{\st}\bigr\}_1^{p-1},\bigl\{\bs^{k}\bigr\}_p^N)\;
T_{p,1}(\bs^1_{\so})\widetilde{\mathbb{B}}(\bt)\\
\times \frac{\prod_{\nu=2}^{p-1}\alpha_\nu(\bs^\nu_{\so})g_{[\nu]}(\bs^{\nu}_{\so},\bs^{\nu-1}_{\so})
 \hg_{\nu}(\bs^{\nu}_{\st},\bs^{\nu}_{\so})}
{\lambda_2(\bs^1_{\so})h(\bs^1,\bs^1_{\so})^{\delta_{m,1}}\prod_{\nu=1}^{p-1}f_{[\nu+1]}(\bs^{\nu+1},\bs^{\nu}_{\so})}.
\end{multline}
The action of $T_{p,1}(\bs^1_{\so})$ onto the vector $\widetilde{\mathbb{B}}(\bt)$
modulus unwanted terms is given by proposition~\ref{P-actTp1}.  Thus, we obtain
\begin{multline}\label{Rec-SP1}
\tilde S(\bs|\bt)\cong \alpha_{1}(\bs^{1}_{\so})\sum_{p=2}^{N+1}
\sum_{\substack{\text{\rm part}(\bs^2,\dots,\bs^{p-1})\\\text{\rm part}(\bt^1,\dots,\bt^{p-1})}} (-1)^{(r_1-1)\delta_{m,1}}
\frac{g_{[2]}(\bt^{1}_{\so},\bs^{1}_{\so})\hg_{1}(\bt^{1}_{\so},\bt^{1}_{\st})
f_{[1]}(\bt^{1}_{\st}, \bs^{1}_{\so} )}{f_{[p]}(\bs^{p},\bs^{p-1}_{\so})h(\bs^1,\bs^1_{\so})^{\delta_{m,1}}}\\
\times\prod_{\nu=2}^{p-1}\frac{\alpha_\nu(\bs^\nu_{\so})g_{[\nu]}(\bs^{\nu}_{\so},\bs^{\nu-1}_{\so})g_{[\nu+1]}(\bt^{\nu}_{\so},\bt^{\nu-1}_{\so})
 \hg_{\nu}(\bs^{\nu}_{\st},\bs^{\nu}_{\so})\hg_{\nu}(\bt^{\nu}_{\so},\bt^{\nu}_{\st})}
{ f_{[\nu]}(\bs^{\nu},\bs^{\nu-1}_{\so})f_{[\nu]}(\bar t^{\nu}_{\so},\bar t^{\nu-1})}\\
\times \mathbb{C}(\bigl\{\bs^k_{\st}\bigr\}_1^{p-1},\bigl\{\bs^{k}\bigr\}_p^N)\;
\widetilde{\mathbb{B}}(\bigl\{\bt^k_{\st}\bigr\}_1^{p-1};\bigl\{\bt^k\bigr\}_p^{N}).
\end{multline}
Here $\bt^{m+n}=\bs^{m+n}=\emptyset$.
Calculating the reduced scalar products in \eqref{Rec-SP1} modulus unwanted terms
\begin{multline}\label{RSP}
\mathbb{C}(\bigl\{\bs^k_{\st}\bigr\}_1^{p-1},\bigl\{\bs^{k}\bigr\}_p^N)
\widetilde{\mathbb{B}}(\bigl\{\bt^k_{\st}\bigr\}_1^{p-1};\bigl\{\bt^k\bigr\}_p^{N})
\cong \prod_{k=1}^{p-1}\alpha_k(\bs^k_{\st})\prod_{\ell=p}^{N}\alpha_\ell(\bs^\ell)\\
\times Z^{m|n}(\bigl\{\bs^k_{\st}\bigr\}_1^{p-1},\bigl\{\bs^{k}\bigr\}_p^N|
\bigl\{\bt^k_{\st}\bigr\}_1^{p-1};\bigl\{\bt^k\bigr\}_p^{N}),
\end{multline}
and substituting this into \eqref{Rec-SP1} we immediately arrive at the recursion \eqref{Rec-HC1}.

 We have also used
$$(-1)^{(r_1-1)\delta_{m,1}}\hg_{1}(\bt^{1}_{\so},\bt^{1}_{\st})=\gamma_{1}(\bt^{1}_{\so},\bt^{1}_{\st}),
\qquad \hg_{\nu}(\bs^{\nu}_{\st},\bs^{\nu}_{\so})\hg_{\nu}(\bt^{\nu}_{\so},\bt^{\nu}_{\st})=
\gamma_{\nu}(\bs^{\nu}_{\st},\bs^{\nu}_{\so})\gamma_{\nu}(\bt^{\nu}_{\so},\bt^{\nu}_{\st}).$$

\subsection{Symmetry of the Highest Coefficient\label{SS-SHC}}

Due to isomorphism \eqref{morphism} between Yangians $Y(\mathfrak{gl}(m|n))$ and $Y(\mathfrak{gl}(n|m))$ one can find a simple
relation between the HC corresponding to these algebras. In this section we obtain this relation.

Consider the sum formula \eqref{al-dep} for the scalar product of $\mathfrak{gl}(m|n)$ Bethe vectors
\begin{equation}\label{al-dep-ord}
S^{m|n}(\ors|\ort)= \sum W^{m|n}_{\text{\rm part}}(\ors_{\!\so},\ors_{\!\st}|\ort_{\!\so},\ort_{\!\st})
  \prod_{k=1}^{N} \alpha_{k}(\bs^k_{\so}) \alpha_{k}(\bt^k_{\st}),
\end{equation}
where we have stressed the ordering \eqref{orders} of the Bethe parameters.
Let us act with the morphism $\varphi$ \eqref{morphism} on the scalar product $S^{m|n}(\ors|\ort)$. This can be done in two ways. First, using
\eqref{morph-BV} and \eqref{morph-dBV} we obtain
\begin{multline}\label{phi-S}
\varphi\Bigl(S^{m|n}(\ors|\ort)\Bigr) =\varphi\Bigl(\mathbb{C}^{m|n}(\ors)\mathbb{B}^{m|n}(\ort)\Bigr)
=\frac{(-1)^{r_m}\mathbb{C}^{n|m}(\ols)\mathbb{B}^{n|m}(\olt)}{\prod_{k=1}^N\alpha_{N+1-k}(\bs^k)\alpha_{N+1-k}(\bt^k)}\\
=\frac{(-1)^{r_m}S^{n|m}(\ols|\olt)}{\prod_{k=1}^N\alpha_{N+1-k}(\bs^k)\alpha_{N+1-k}(\bt^k)}.
\end{multline}
The scalar product $S^{n|m}(\ols|\olt)$ has the standard representation \eqref{al-dep}. Thus, we find
\begin{equation}\label{1-way}
\varphi\Bigl(S^{m|n}(\ors|\ort)\Bigr)= \sum_{\text{part}}
\frac{(-1)^{r_m}W^{n|m}_{\text{\rm part}}(\ols_{\!\so},\ols_{\!\st}|\olt_{\!\so},\olt_{\!\st})}
{\prod_{k=1}^N\alpha_{N+1-k}(\bs^k)\alpha_{N+1-k}(\bt^k)}  \prod_{k=1}^{N} \alpha_{k}(\bs^{N-k+1}_{\so}) \alpha_{k}(\bt^{N-k+1}_{\st}).
\end{equation}

On the other hand, acting with $\varphi$ directly on the sum formula \eqref{al-dep-ord} we have
\be{SP-part}
\varphi\Bigl(S^{m|n}(\ors|\ort)\Bigr)=\sum_{\text{part}}
 W^{m|n}_{\text{part}}(\ors_{\!\so},\ors_{\!\st}|\ort_{\!\so},\ort_{\!\st})
\prod_{k=1}^{N}\Bigl(\alpha_{N+1-k}(\bs^k_{\so})\alpha_{N+1-k}(\bt^k_{\st})\Bigr)^{-1}.
\ee
Comparing \eqref{1-way} and \eqref{SP-part} we arrive at
\begin{multline}\label{compare}
 (-1)^{r_m}\sum_{\text{part}}
W^{n|m}_{\text{\rm part}}(\ols_{\!\so},\ols_{\!\st}|\olt_{\!\so},\olt_{\!\st})
 \prod_{k=1}^{N}\alpha_{N+1-k}(\bs^k_{\so})\alpha_{N+1-k}(\bt^k_{\st})\\
=\sum_{\text{part}}
W^{m|n}_{\text{part}}(\ors_{\!\so},\ors_{\!\st}|\ort_{\!\so},\ort_{\!\st})
\prod_{k=1}^{N}\alpha_{N+1-k}(\bs^k_{\st})\alpha_{N+1-k}(\bt^k_{\so})
\end{multline}
Since $\alpha_i$ are free functional parameters, the coefficients of the same products of $\alpha_i$ must be equal.
Hence,
\be{W-W}
W^{m|n}_{\text{part}}(\ors_{\!\so},\ors_{\!\st}|\ort_{\!\so},\ort_{\!\st})=
(-1)^{r_m}W^{n|m}_{\text{\rm part}}(\ols_{\!\st},\ols_{\!\so}|\olt_{\!\st},\olt_{\!\so}),
\ee
for arbitrary partitions of the sets $\bs$ and $\bt$. In particular, setting $\bs_{\st}=\bt_{\st}=\emptyset$ we obtain
\be{Z-Z}
Z^{m|n}(\ors|\ort)=(-1)^{r_m}\overline{Z}^{n|m}(\ols|\olt)=(-1)^{r_m}Z^{n|m}(\olt|\ols).
\ee
Using this property one can obtain  recursion \eqref{Rec-2HC1} for the highest coefficient. Indeed, one can easily see that applying \eqref{Rec-HC1}
to the rhs of \eqref{Z-Z} we obtain \eqref{Rec-2HC1} for the lhs of this equation.

\section*{Conclusion}

In the present paper we have considered the Bethe vectors scalar products in the integrable models solvable by the nested
algebraic Bethe ansatz and possessing $\mathfrak{gl}(m|n)$ supersymmetry. The main result of the paper is the sum formula given by equations \eqref{al-dep} and \eqref{HypResh}. We obtained it using the coproduct formula for the Bethe vectors. This way certainly is more direct and simple than the methods used
before for the derivation of the sum formulas.

The sum formula is obtained for the Bethe vectors with arbitrary coloring. However, as we have mentioned in section~\ref{SS-C},
in various models of physical interest the coloring of the Bethe vectors is restricted by the condition
$r_1\ge r_2\ge\dots\ge r_N$. A peculiarity of these models is that only the ratio $\alpha_1(u)$ is a non-trivial function of $u$, while
all other $\alpha$'s are identically constants: $\alpha_k(u)=\alpha_k$, $k>1$ (actually, using a twist transformation, one can always
make these constants equal to $1$: $\alpha_k(u)=1$, $k>1$). Then equation \eqref{al-dep} is simplified, and one can try to take the
sum over most of partitions, what should lead to a significant simplification of the sum formula. This direction of possible development is very attractive, and we are planning to study this problem.

The sum formula involves the HC of the scalar product. We did not find a closed expression for the HC, however, we have found recursions for it.
Perhaps, this way of describing the HC is preferable for the models with high rank of symmetry. Indeed, looking at the explicit formulas for the HC
in the $\mathfrak{gl}(3)$-based models one hardly can expect to obtain a relatively  simple closed formula for it in the general $\mathfrak{gl}(m|n)$
case. On the other hand, the recursions obtained in this paper allow one to study analytical properties of the HC, in particular to find the residues
in the poles of this rational function.  Using these results it is possible to derive an analog of Gaudin formula for on-shell Bethe vectors
in the $\mathfrak{gl}(m|n)$ based models exactly in the same way as it was done in \cite{Kor82,Res86}. We will consider this question in our
forthcoming publication.

As we have already mentioned  in Introduction, the sum formula itself is not  very convenient for use. One should remember,
however, that the sum formula describes the scalar product of generic Bethe vectors, where we have no restriction for the Bethe parameters.
At the same time, in most cases of physical interest one deals with Bethe vectors, in which most of the Bethe parameters satisfy Bethe equations.
In particular, this situation occurs in calculating form factors. Then one can hope to obtain a significant simplification of the sum formula, as
it was shown for the models with $\mathfrak{gl}(3)$ and $\mathfrak{gl}(2|1)$ symmetries. We are planning to study this problem in our further publications.

In conclusion we would like to discuss one more possible direction of generalization of our results. In this paper we considered
the so-called distinguished gradation, that is to say the special
grading $[i]=0$ for $1\le i\le m$, $[i]=1$ for $m<i\le m+n$. However, this is not the only possible choice of grading.
Other gradings induce different inequivalent presentations of the superalgebra, where the number of fermionic simple roots can vary from a presentation to another. These different presentations are labelled by the different Dynkin diagrams associated to the superalgebra. Obviously, since the different presentations deal with the same superalgebra, they are isomorphic. However, the mapping between two presentations is  based on
a generalized Weyl transformation acting on their Dynkin diagrams, lifted at the level of the superalgebra. These generalized Weyl transformations,
in particular, affect the bosonic/fermionic nature of the generators, and thus can change commutators to anti-commutators (and vice-versa). Then,
the precise expression of the mapping
is heavy to formulate for all the generators of the Yangian. This is also true for Bethe vectors and Bethe parameters, a precise correspondence can be quite intricate to formulate.
However, from the Lie superalgebra theory one knows that such a correspondence must exist. These considerations have been developed  in \cite{Frank} for the construction of the mapping on the particular case of the
$\mathfrak{gl}(1|2)$ algebra. The general case of generic $\mathfrak{gl}(m|n)$ superalgebra is presented in \cite{AACRFE} for the form of the Bethe equations,
but open spin chains (see also \cite{Satta} where the periodic case is reviewed). In conclusion, if a qualitative
 generalization of the present results to
the superalgebras with different gradings is rather straightforward, a precise correspondence remains open.

\section*{Acknowledgements}
The work of A.L. has been funded by Russian Academic Excellence Project 5-100, by Young Russian Mathematics award and by joint NASU-CNRS project F14-2017.
The work of S.P. was supported in part by the RFBR grant 16-01-00562-a.


\appendix

\section{Coproduct formula for the Bethe vectors\label{SS-CPFSP}}

The presentation \eqref{BB11} for the Bethe vector of the composite model can be treated as a coproduct formula for the
Bethe vector. Indeed, equation \eqref{T-TT} formally determines a coproduct $\Delta$ of the monodromy matrix entries
\begin{equation}\label{cop1}
  \Delta(T_{i,j}(u)) = \sum_{k=1}^{m+n} (-1)^{([j]+[k])([i]+[k])}T_{k,j}(u)\otimes T_{i,k}(u).
\end{equation}

Then \eqref{BB11} is nothing but the action of $\Delta$ onto the Bethe vector \cite{HutLPRS17a}.

The action of the coproduct onto the dual Bethe vectors can be obtained via antimorphism \eqref{antimo}. It was proved in
 \cite{PakRS15d}  (see also similar consideration in prop. 1.5.4 of \cite{alexbook})  that
\begin{equation}\label{cop3}
  \Delta \circ \Psi  = (\Psi \otimes \Psi) \circ \Delta',
\end{equation}
where
\begin{equation}\label{cop2}
  \Delta'(T_{i,j}(u)) = \sum T_{i,k}(u)\otimes T_{k,j}(u).
\end{equation}
Then
\begin{multline}\label{cop4}
  \Delta(\mathbb{C}(\bar t)) = \Delta(\Psi(\mathbb{B}(\bar t))) =
  (\Psi\otimes\Psi) \circ \Delta'( \mathbb{B}(\bar t) ) \\
=(\Psi\otimes\Psi)
 \left( \sum\frac{
 \prod_ {\nu=1}^{N} \alpha^{(1)}_{\nu}(\bt^\nu_{\so}) \gamma_{\nu}(\bt^\nu_{\st},\bt^\nu_{\so})}
   { \prod_{\nu=1}^{N-1} f_{[\nu+1]}(\bt^{\nu+1}_{\st},\bt^\nu_{\so})}\;
 \mathbb{B}^{(2)}(\bt_{\so})  \otimes
 \mathbb{B}^{(1)}(\bt_{\st}) \right)\\
 =\sum\frac{
 \prod_ {\nu=1}^{N}  \alpha^{(1)}_{\nu}(\bt^\nu_{\so})\gamma_{\nu}(\bt^\nu_{\st},\bt^\nu_{\so})}
{\prod_{\nu=1}^{N-1} f_{[\nu+1]}(\bt^{\nu+1}_{\st},\bt^\nu_{\so})   }\;
 \mathbb{C}^{(2)}(\bt_{\so}) \otimes \mathbb{C}^{(1)}(\bt_{\st}).
\end{multline}
Relabeling here the subsets  $\bt^\nu_{\so} \leftrightarrow  \bt^\nu_{\st}$ we arrive at \eqref{BC1}. \qed

\section{Action formulas}

In this section we derive the action of the operators $T_{p,1}$ on the main term \eqref{Norm-MT}. For this
we first consider some multiple commutation relations in the $RTT$-algebra \eqref{RTT}.

\subsection{Multiple commutation relations}

Multiple commutation relations of the monodromy matrix entries in superalgebras were stu\-di\-ed in \cite{Sla16}. Here we consider
several particular cases of commutation relations with the operators $\bT_{i,i+1}(\bv)$ \eqref{bT}.

It follows  from \eqref{TM-1} that
\be{sing-CR}
\begin{aligned}
T_{i,i}(u)T_{i,i+1}(v)&= f_{[i]}(v,u)T_{i,i+1}(v)T_{i,i}(u) + g_{[i]}(u,v)T_{i,i+1}(u)T_{i,i}(v),\\
T_{i,i}(u)T_{i-1,i}(v)&= f_{[i]}(u,v)T_{i-1,i}(v)T_{i,i}(u) + g_{[i]}(v,u)T_{i-1,i}(u)T_{i,i}(v).
\end{aligned}
\ee
We see that these commutation relations look exactly the same as in the case of algebra $\mathfrak{gl}(n)$.
The only difference is that the functions $f$ and $g$ acquire an additional subscript  indicating parity.
Therefore, for commutation relations, we can apply the standard arguments of the algebraic Bethe ansatz \cite{FadST79,BogIK93L,FadLH96}.
In particular, let us consider commutation of the operator $T_{i,i}(t^{i-1}_\alpha)$ with the product $\bT_{i,i+1}(\bt^{i})$, where
$t^{i-1}_\alpha$ is a fixed parameter of the set $\bt^{i-1}$. Let us call a term wanted, if it contains the operator $T_{i,i}(t^{i-1}_\alpha)$
in the extreme right position. Then moving $T_{i,i}(t^{i-1}_\alpha)$ through the product $\bT_{i,i+1}(\bt^{i})$
we should keep the original argument of $T_{i,i}$ leading to
\be{Tii-bTii1}
T_{i,i}(t^{i-1}_\alpha)\bT_{i,i+1}(\bt^{i})\cong f_{[i]}(\bt^{i},t^{i-1}_\alpha) \bT_{i,i+1}(\bt^{i})T_{i,i}(t^{i-1}_\alpha).
\ee
%
%
%
%
%

Consider now commutation of the operator $T_{i+1,i}(t^{i-1}_\alpha)$ with the product $\bT_{i,i+1}(\bt^{i})$ using
\begin{multline}\label{sing-CR2}
T_{i+1,i}(u)T_{i,i+1}(v)-(-1)^{\delta_{i,m}} T_{i,i+1}(v)T_{i+1,i}(u)\\
=g_{[i+1]}(u,v)\bigl(T_{i+1,i+1}(u)T_{i,i}(v)-T_{i+1,i+1}(v)T_{i,i}(u)\bigr).
\end{multline}
Let, as before, a term be wanted, if it contains the operator $T_{i,i}(t^{i-1}_\alpha)$
in the extreme right position.
Moving  $T_{i+1,i}(t^{i-1}_\alpha)$ through the product $\bT_{i,i+1}(\bt^{i})$ we can obtain the terms of the
following type:
\be{types10}
\begin{aligned}
&({\rm i})\quad T_{i+1,i}(t^{i-1}_\alpha);\\
&({\rm ii})\quad T_{i+1,i+1}(t^{i}_j)T_{i,i}(t^{i-1}_\alpha),\qquad j=1,\dots,r_i;\\
&({\rm iii})\quad T_{i+1,i+1}(t^{i-1}_\alpha)T_{i,i}(t^{i}_j), \qquad j=1,\dots,r_i;\\
&({\rm iv})\quad T_{i+1,i+1}(t^{i}_{j_1})T_{i,i}(t^{i}_{j_2}), \qquad j_1,j_2=1,\dots,r_i.
\end{aligned}
\ee
Among all these contributions only the terms (ii) are wanted. Thus, we have
\be{prelim-res}
T_{i+1,i}(t^{i-1}_\alpha)\bT_{i,i+1}(\bt^{i})\cong \sum_{j=1}^{r_i} \Lambda_j
\bT_{i,i+1}(\bt^{i}\setminus t^{i}_j)T_{i+1,i+1}(t^{i}_j)T_{i,i}(t^{i-1}_\alpha),
\ee
where $\Lambda_j $ are rational coefficients to be determined. Due to the symmetry of $\bT_{i,i+1}(\bt^{i})$ over
$\bt^{i}$ it is sufficient to find $\Lambda_1 $ only. Then a wanted term must contain $T_{i+1,i+1}(t^{i}_1)T_{i,i}(t^{i-1}_\alpha)$
in the extreme right position. We have
\begin{multline}\label{step1}
T_{i+1,i}(t^{i-1}_\alpha)\bT_{i,i+1}(\bt^{i})= T_{i+1,i}(t^{i-1}_\alpha)\frac{T_{i,i+1}(t^{i}_1)\bT_{i,i+1}(\bt^{i}\setminus t^{i}_1)}
{h(\bt^{i},t^{i}_1)^{\delta_{m,i}} }\\
\cong g_{[i+1]}(t^{i-1}_\alpha,t^{i}_1)\bigl(T_{i+1,i+1}(t^{i-1}_\alpha)  T_{i,i}(t^{i}_1) -T_{i+1,i+1}(t^{i}_1)T_{i,i}(t^{i-1}_\alpha)\bigr)
\frac{\bT_{i,i+1}(\bt^{i}\setminus t^{i}_1)}{h(\bt^{i},t^{i}_1)^{\delta_{m,i}} }.
\end{multline}
The term $T_{i+1,i+1}(t^{i-1}_\alpha)  T_{i,i}(t^{i}_1) $ obviously gives unwanted contribution. The remaining operators
$T_{i+1,i+1}(t^{i}_1)T_{i,i}(t^{i-1}_\alpha)$ should move through the product $\bT_{i,i+1}(\bt^{i}\setminus t^{i}_1)$ via
\eqref{sing-CR} keeping their arguments. This leads to
\begin{multline}\label{step12}
T_{i+1,i}(t^{i-1}_\alpha)\bT_{i,i+1}(\bt^{i})
\cong g_{[i+1]}(t^{i}_1,t^{i-1}_\alpha) \prod_{k=2}^{r_i} f_{[i]}(t^{i}_k,t^{i-1}_\alpha)  f_{[i+1]}(t^{i}_1,t^{i}_k)\\
\times
\frac{\bT_{i,i+1}(\bt^{i}\setminus t^{i}_1)}{h(\bt^{i},t^{i}_1)^{\delta_{m,i}} }T_{i+1,i+1}(t^{i}_1)T_{i,i}(t^{i-1}_\alpha).
\end{multline}
Thus, using \eqref{gam-hg} we arrive at
\be{Lam1}
\Lambda_1=g_{[i+1]}(t^{i}_1,t^{i-1}_\alpha) \prod_{k=2}^{r_i} f_{[i]}(t^{i}_k,t^{i-1}_\alpha)  \hg_{i}(t^{i}_1,t^{i}_k).
\ee

The final result can be written as a sum over partitions of the set $\bt^{i}$:
\begin{multline}\label{Ti1i-bTii1}
T_{i+1,i}(t^{i-1}_\alpha)\bT_{i,i+1}(\bt^{i})
\cong \sum g_{[i+1]}(\bt^{i}_{\so},t^{i-1}_\alpha)
f_{[i]}(\bt^{i}_{\st},t^{i-1}_\alpha)\hg_{i}(\bt^{i}_{\so},\bt^{i}_{\st})\\
\times\bT_{i,i+1}(\bt^{i}_{\st})\;T_{i+1,i+1}(\bt^{i}_{\so})T_{i,i}(t^{i-1}_\alpha).
\end{multline}
Here the set $\bt^{i}$ is divided into subsets $\bt^{i}_{\so}$ and $\bt^{i}_{\st}$ such that $\#\bt^{i}_{\so}=1.$

\subsection{Action formulas}

In this section we consider the action of the operators  $T_{p,1}(s)$ onto the  main term of the Bethe vector \eqref{Norm-MT}. Here
$p>1$ and $s$ is a generic complex number. The result of this action contains various terms, among which we will distinguish wanted
and unwanted terms. Let a term be wanted, if it is proportional to $\lambda_{1}(s)$  and does not contain any $\alpha_i(t^k_\ell)$. Otherwise a term is unwanted.

\begin{prop}\label{P-actTp1}
Let $\widetilde{\mathbb{B}}(\bt)$ be the main term of a Bethe vector \eqref{Norm-MT}. Then the wanted term of the action of $T_{p,1}$
onto $\widetilde{\mathbb{B}}(\bt)$ reads
\begin{multline}\label{Tp1-tB}
T_{p,1}(s)\widetilde{\mathbb{B}}(\bt)
              \cong \lambda_{1}(s)\sum_{\text{\rm part}(\bt)}
 \prod_{\ell=2}^{p-1}
\frac{g_{[\ell+1]}(\bt^{\ell}_{\so},\bt^{\ell-1}_{\so})
\hg_{\ell}(\bt^{\ell}_{\so},\bt^{\ell}_{\st})}
{ f_{[\ell]}(\bt^{\ell}_{\so},\bt^{\ell-1})}\\
\times g_{[2]}(\bt^{1}_{\so},s)\hg_{1}(\bt^{1}_{\so},\bt^{1}_{\st})
f_{[1]}(\bt^{1}_{\st},s)\widetilde{\mathbb{B}}(\bigl\{\bt^k_{\st}\bigr\}_1^{p-1};\bigl\{\bt^k\bigr\}_p^{N}).
\end{multline}
Here the sum is taken over partitions of the sets $\bt^k$ with $k=1,\dots, p-1$ into subsets $\bt^k_{\so}$ and $\bt^k_{\st}$
such that $\#\bt^k_{\so}=1$.
\end{prop}

To prove proposition~\ref{P-actTp1} we introduce for $1\le i<k\le m+n$
\be{phi}
\widetilde{\mathbb{B}}_{ik}(\{\bt^\nu\}_{i}^{k-1})=
\frac{\bT_{i,i+1}(\bt^i)\dots \bT_{k-1,k}(\bt^{k-1})|0\rangle}{\prod_{j=i}^{k-1}\lambda_{j+1}(\bt^j)
\prod_{j=i}^{k-2}f_{[j+1]}(\bt^{j+1},\bt^j)},
\ee
where $\bT_{j,j+1}$ is defined by \eqref{bT}. Obviously, $\widetilde{\mathbb{B}}_{1,n+m}(\{\bt^\nu\}_{1}^{N})=\widetilde{\mathbb{B}}(\bt)$.
We first prove several auxiliary lemmas.

\begin{lemma}\label{L1}
Let $j<\ell$ and $j<i$. Then   
\be{Tji-pji}
T_{\ell,j}(s)\widetilde{\mathbb{B}}_{ik}(\{\bt^\nu\}_{i}^{k-1})=0.
\ee
\end{lemma}
{\sl Proof.} The proof is based on the arguments of the coloring. The operator $T_{\ell,j}$ annihilates the particles of
the colors $j,\dots,\ell-1$. On the other hand, for $i>j$ the state $\widetilde{\mathbb{B}}_{ik}(\{\bt^\nu\}_{i}^{k-1})$
does not contain the particles of the color $j$. Hence, the action of $T_{\ell,j}$ onto $\widetilde{\mathbb{B}}_{ik}(\{\bt^\nu\}_{i}^{k-1})$
vanishes. \qed

\begin{lemma}\label{L2}
Let  $j<i$. Then
\be{Tii-pkn}
T_{j,j}(s)\widetilde{\mathbb{B}}_{ik}(\{\bt^\nu\}_{i}^{k-1})=\lambda_{j}(s)\widetilde{\mathbb{B}}_{ik}(\{\bt^\nu\}_{i}^{k-1}).
\ee
\end{lemma}
{\sl Proof.}  Obviously,
\be{present}
\widetilde{\mathbb{B}}_{ik}(\{\bt^\nu\}_{i}^{k-1})= \frac{\bT_{i,i+1}(\bt^{i})}
{\lambda_{i+1}(\bt^i)f_{[i+1]}(\bt^{i+1},\bt^{i})}
\widetilde{\mathbb{B}}_{i+1,k}(\{\bt^\nu\}_{i+1}^{k-1}).
\ee
 When one commutes $T_{j,j}$ with one of the operators in the product $\bT_{i,i+1}(\bt^{i})$, then from \eqref{TM-1},
 we obtain the operators
$T_{i, j}$ or $T_{i+1, j}$ acting on $\widetilde{\mathbb{B}}_{i+1,k}(\bt)$. Due to lemma~\ref{L1} this action vanishes, because $i>j$.
Thus,
\be{Tii-pkn01}
T_{j,j}(s)\widetilde{\mathbb{B}}_{ik}(\{\bt^\nu\}_{i}^{k-1})=\frac{\bT_{i,i+1}(\bt^{i})}
{\lambda_{i+1}(\bt^i)f_{[i+1]}(\bt^{i+1},\bt^{i})}
T_{j,j}(s)\widetilde{\mathbb{B}}_{i+1,k}(\{\bt^\nu\}_{i+1}^{k-1}).
\ee
Continuing this process we eventually move $T_ {j,j}$ to the vacuum vector, where it gives  $\lambda_{j}(s)$. \qed

In the following lemmas the actions are considered modulus unwanted terms. Let $t^{i-1}_\alpha$ be a fixed parameter of the set $\bt^{i-1}$. We say that a term is {\it wanted}, if a Bethe parameter $t_\ell^j$ for $j=i,\dots,k-1$ becomes an argument of $\lambda_{j+1}$. Otherwise, a term is {\it unwanted}.

\begin{lemma}\label{L3} The wanted term of the action of $T_{i,i}(t^{i-1}_\alpha)$ onto $\widetilde{\mathbb{B}}_{ik}(\{\bt^\nu\}_{i}^{k-1})$ is given by
\be{Tkk-pkn}
T_{i,i}(t^{i-1}_\alpha)\widetilde{\mathbb{B}}_{ik}(\{\bt^\nu\}_{i}^{k-1})\cong\lambda_{i}(t^{i-1}_\alpha)
f_{[i]}(\bt^{i},t^{i-1}_\alpha)\widetilde{\mathbb{B}}_{ik}(\{\bt^\nu\}_{i}^{k-1}).
\ee
\end{lemma}
{\sl Proof.} We present $\widetilde{\mathbb{B}}_{ik}(\{\bt^\nu\}_{i}^{k-1})$ in the form \eqref{present}. Then,
moving  $T_{i,i}(t^{i-1}_\alpha)$ through the product $\bT_{i,i+1}(\bt^{i})$ we should use \eqref{Tii-bTii1},
otherwise we obtain unwanted terms. Therefore, at the first step we obtain
\be{Tkk-pkn1}
T_{i,i}(t^{i-1}_\alpha)\widetilde{\mathbb{B}}_{ik}(\{\bt^\nu\}_{i}^{k-1})\cong \frac{f_{[i]}(\bt^{i},t^{i-1}_\alpha)\bT_{i,i+1}(\bt^{i})}
{\lambda_{i+1}(\bt^i)f_{[i+1]}(\bt^{i+1},\bt^{i})} T_{i,i}(t^{i-1}_\alpha)
\widetilde{\mathbb{B}}_{i+1,k}(\{\bt^\nu\}_{i+1}^{k-1}).
\ee
Then application of lemma~\ref{L2} completes the proof. \qed

\begin{lemma}\label{L4}
The wanted term of the action of $T_{i+1,i}(t^{i-1}_\alpha)$ onto $\widetilde{\mathbb{B}}_{ik}(\{\bt^\nu\}_{i}^{k-1})$ is given by
\begin{equation}\label{Tk1k-pkn-p}
T_{i+1,i}(t^{i-1}_\alpha)\widetilde{\mathbb{B}}_{ik}(\{\bt^\nu\}_{i}^{k-1})\cong\sum \lambda_{i}(t^{i-1}_{\alpha})
g_{[i+1]}(\bt^{i}_{\so},t^{i-1}_\alpha)
f_{[i]}(\bt^{i}_{\st},t^{i-1}_\alpha)\hg_{i}(\bt^{i}_{\so},\bt^{i}_{\st})
\widetilde{\mathbb{B}}_{ik}(\bt^{i}_{\st};\{\bt^\nu\}_{i+1}^{k-1}).
\end{equation}
Here the sum is taken over partitions $\bt^{i}\Rightarrow\{\bt^{i}_{\so},\bt^{i}_{\st}\}$ such that $\#\bt^{i}_{\so}=1$.
\end{lemma}
{\sl Proof.}  We again present $\widetilde{\mathbb{B}}_{ik}(\{\bt^\nu\}_{i}^{k-1})$ in the form \eqref{present}. Then,
moving  $T_{i+1,i}(t^{i-1}_\alpha)$ through the product $\bT_{i,i+1}(\bt^{i})$ we should use \eqref{Ti1i-bTii1},
otherwise we obtain unwanted terms. Thus, we obtain
\begin{multline}\label{Tk1k-pkn1}
T_{i+1,i}(t^{i-1}_\alpha)\widetilde{\mathbb{B}}_{i+1,k}(\bt)\cong
\sum g_{[i+1]}(\bt^{i}_{\so},t^{i-1}_\alpha)
f_{[i]}( \bt^{i}_{\st},t^{i-1}_\alpha)\hg_{i}(\bt^{i}_{\so},\bt^{i}_{\st})\\
\times
\frac{\bT_{i,i+1}(\bt^{i}_{\st})\;T_{i+1,i+1}(\bt^{i}_{\so})T_{i,i}(t^{i-1}_\alpha)}{\lambda_{i+1}(\bt^i)f_{[i+1]}(\bt^{i+1},\bt^{i})}\;
\widetilde{\mathbb{B}}_{i+1,k}(\{\bt^\nu\}_{i+1}^{k-1}).
\end{multline}
Then application of lemmas~\ref{L2}  and~\ref{L3} completes the proof. \qed

\begin{lemma}\label{L5}
Let $i<p< k$. Then
\begin{multline}\label{Tpk-pkn}
T_{p,i}(t^{i-1}_\alpha)\widetilde{\mathbb{B}}_{ik}(\{\bt^\nu\}_{i}^{k-1})
\cong \lambda_{i}(t^{i-1}_\alpha)\sum_{\text{\rm part}(\bt)}
\widetilde{\mathbb{B}}_{ik}(\{\bt^{\nu}_{\st}\}_i^{p-1};\{\bt^\nu\}_{p}^{k-1})\\
\times
g_{[i+1]}(\bt^{i}_{\so},t^{i-1}_\alpha)\hg_{i}(\bt^{i}_{\so},\bt^{i}_{\st})f_{[i]}(\bt^{i}_{\st},t^{i-1}_\alpha)
\prod_{\nu=i+1}^{p-1}\frac{g_{[\nu+1]}(\bt^{\nu}_{\so},\bt^{\nu-1}_{\so})\hg_{\nu}(\bt^{\nu}_{\so},\bt^{\nu}_{\st})}
{f_{[\nu]}(\bt^{\nu}_{\so},\bt^{\nu-1})}.
\end{multline}
Here the sum is taken over partitions of the sets $\bt^{\nu}\Rightarrow\{\bt^{\nu}_{\so},\bt^{\nu}_{\st}\}$ for $\nu=i,\dots,p-1$, such
that $\#\bt^{\nu}_{\so}=1$.
\end{lemma}
{\sl Proof.} The proof uses induction over $p-i$. If $p-i=1$, then the statement  coincides with the one of lemma~\ref{L4}. Assume
that \eqref{Tpk-pkn} is valid for $i$ replaced with $i+1$. Then we use presentation \eqref{present}
\be{shag1}
T_{p,i}(t^{i-1}_{\alpha})\widetilde{\mathbb{B}}_{ik}(\{\bt^\nu\}_{i}^{k-1})=
\frac{T_{p,i}(t^{i-1}_{\alpha})\bT_{i,i+1}(\bt^{i})}
{\lambda_{i+1}(\bt^i)f_{[i+1]}(\bt^{i+1},\bt^{i})}
\widetilde{\mathbb{B}}_{i+1,k}(\{\bt^\nu\}_{i+1}^{k-1}).
\ee
Moving  $T_{p,i}(t^{i-1}_{\alpha})$ through the product $\bT_{i,i+1}(\bt^{i})$ we can obtain the terms of the
following type:
\be{types2}
\begin{aligned}
&({\rm i})\quad T_{p,i}(t^{i-1}_{\alpha});\\
&({\rm ii})\quad T_{p,i+1}(t^{i}_j)T_{i,i}(t^{i-1}_{\alpha});\\
&({\rm iii})\quad T_{p,i+1}(t^{i-1}_{\alpha})T_{i,i}(t^{i}_j);\\
&({\rm iv})\quad T_{p,i+1}(t^{i}_{j_1})T_{i,i}(t^{i}_{j_2}).
\end{aligned}
\ee
The term (i) vanishes due to lemma~\ref{L1}. The terms (iii) and (iv) give unwanted
terms due to lemma~\ref{L2}. Hence, only the term (ii) survives. Using the arguments similar to the ones that we used for obtaining
equation \eqref{Ti1i-bTii1} we arrive at
\begin{multline}\label{Tpk-pkn1}
T_{p,i}(t^{i-1}_{\alpha})\widetilde{\mathbb{B}}_{ik}(\{\bt^\nu\}_{i}^{k-1})\cong
\sum g_{[i+1]}(\bt^{i}_{\so},t^{i-1}_\alpha)
f_{[i]}(\bt^{i}_{\st},t^{i-1}_\alpha)\hg_{i}(\bt^{i}_{\so},\bt^{i}_{\st})\\
\times
\frac{\bT_{i,i+1}(\bt^{i}_{\st})\;T_{p,i+1}(\bt^{i}_{\so})T_{i,i}(t^{i-1}_\alpha)}{\lambda_{i+1}(\bt^i)f_{[i+1]}(\bt^{i+1},\bt^{i})}\;
\widetilde{\mathbb{B}}_{i+1,k}(\{\bt^\nu\}_{i+1}^{k-1}).
\end{multline}
Here the sum is taken over partitions $\bt^{i}\Rightarrow\{\bt^{i}_{\so},\bt^{i}_{\st}\}$ such that $\#\bt^{i}_{\so}=1$.
Applying  lemma~\ref{L2}  we find
\begin{multline}\label{Tpk-pkn3}
T_{p,i}(t^{i-1}_{\alpha})\widetilde{\mathbb{B}}_{ik}(\{\bt^\nu\}_{i}^{k-1})\cong \sum \lambda_{i}(t^{i-1}_{\alpha})g_{[i+1]}(\bt^{i}_{\so},t^{i-1}_{\alpha})
f_{[i]}(\bt^{i}_{\st}, t^{i-1}_{\alpha})\hg_{i}(\bt^{i}_{\so},\bt^{i}_{\st})\\
\times
\frac{\bT_{i,i+1}(\bt^{i}_{\st})\;T_{p,i+1}(\bt^{i}_{\so})}{\lambda_{i+1}(\bt^i)f_{[i+1]}(\bt^{i+1},\bt^{i})}\;
\widetilde{\mathbb{B}}_{i+1,k}(\{\bt^\nu\}_{i+1}^{k-1}).
\end{multline}
 The action
of $T_{p,i+1}(\bt^{i}_{\so})$ onto $\widetilde{\mathbb{B}}_{i+1,k}(\{\bt^\nu\}_{i+1}^{k-1})$ is known due to the induction assumption. Substituting this
known action into \eqref{Tpk-pkn1} we prove lemma~\ref{L5}. \qed

In fact, lemma~\ref{L5} gives the proof of proposition~\ref{P-actTp1}. Indeed, it is enough to set $i=1$ and $k=m+n$ in \eqref{Tpk-pkn}. We also
 set by definition $t^{0}_{\alpha}=s$ and  introduce an auxiliary empty
set $\bt^{m+n}\equiv \emptyset$. Then lemma~\ref{L5} describes the action of $T_{p,1}(s)$ onto the main term $\widetilde{\mathbb{B}}(\bt)$.


\begin{thebibliography}{99}
%
\bibitem{FadST79} L. D. Faddeev, E. K. Sklyanin and L. A. Takhtajan, \textsl{Quantum Inverse Problem. I},
 Theor. Math. Phys. {\bf 40} (1979) 688--706.
 %
\bibitem{FadT79} L. D. Faddeev and L. A. Takhtajan, {\sl The quantum method of the inverse problem and the Heisenberg $XYZ$ model},
Usp. Math. Nauk {\bf 34} (1979) 13;  Russian Math. Surveys {\bf 34} (1979) 11 (Engl. transl.).
%
\bibitem{BogIK93L}V. E. Korepin, N. M. Bogoliubov,
A. G. Izergin, \textsl{Quantum Inverse Scattering Method and Correlation Functions}, Cambridge: Cambridge Univ.
Press, 1993.
%
\bibitem{FadLH96} L. D. Faddeev, in: Les Houches Lectures \textsl{Quantum Symmetries}, eds A. Connes
et al, North Holland, (1998) 149.
%
\bibitem{Kor82} V. E. Korepin, \textsl{Calculation of norms of Bethe wave functions}, Comm. Math. Phys. {\bf 86} (1982) 391--418.
%
\bibitem{IzeK84}
 A. G. Izergin and V. E. Korepin, \textsl{The Quantum Inverse Scattering Method Approach to
Correlation Functions}, Commun. Math. Phys. {\bf 94} (1984) 67--92.
%
 \bibitem{KulRes83}
P. P. Kulish, N. Yu. Reshetikhin,
{\sl Diagonalization of $GL(N)$ invariant transfer matrices and quantum $N$-wave system (Lee model)}, J.~Phys.~A:  {\bf 16} (1983) L591--L596.
%
 \bibitem{KulRes81}
P. P. Kulish, N. Yu. Reshetikhin,
{\sl Generalized Heisenberg ferromagnet and the Gross--Neveu model}, Zh. Eksp. Theor. Fiz.
{\bf 80} (1981) 214--228; Sov. Phys. JETP,  {\bf 53}:1 (1981)  108--114 (Engl. transl.)
%
 \bibitem{KulRes82}
P. P. Kulish, N. Yu. Reshetikhin,
{\sl GL(3)-invariant solutions of the Yang-Baxter equation and associated quantum systems},
Zap. Nauchn. Sem. LOMI. {\bf 120} (1982) 92--121; J. Sov. Math.,  {\bf 34}:5 (1986)  1948--1971 (Engl. transl.)
%
\bibitem{Res86}
N. Yu. Reshetikhin, {\sl Calculation of the norm of Bethe vectors in models with $SU(3)$-symmetry},
Zap. Nauchn. Sem. LOMI {\bf 150} (1986) 196--213;    J. Math. Sci. {\bf 46} (1989) 1694--1706 (Engl. transl.).
%
\bibitem{Whe12} M. Wheeler, \textsl{Scalar products in generalized models with $SU(3)$-symmetry}, Comm. Math. Phys. {\bf 327}:3
(2014) 737--777, \texttt{arXiv:1204.2089}.
%
\bibitem{Whe13} M. Wheeler, \textsl{Multiple integral formulae for the scalar product of on-shell and off-shell Bethe vectors
in $SU(3)$-invariant models}, Nucl. Phys. B {\bf 875}:1 (2013) 186--212, \texttt{arXiv:1306.0552}.
%
\bibitem{BelPRS12a} S. Belliard, S. Pakuliak, E. Ragoucy, N. A. Slavnov,
\textsl{Highest coefficient of scalar products in $SU(3)$-invariant integrable models},
J. Stat. Mech. Theory Exp., (2012) P09003, \texttt{arXiv:1206.4931}.
%
\bibitem{BelPRS12b} S. Belliard, S. Pakuliak, E. Ragoucy, N. A. Slavnov,
\textsl{The algebraic Bethe ansatz for scalar products in $SU(3)$-invariant integrable
 models}, J. Stat. Mech. Theory Exp., (2012) P10017, \texttt{arXiv:1207.0956}.
%
\bibitem{HutLPRS16c} A. Hutsalyuk, A. Liashyk, S. Pakuliak, E. Ragoucy, N. A. Slavnov, \textsl{
Scalar products of Bethe vectors in models with $\mathfrak{gl}(2|1)$ symmetry 1. Super-analog of Reshetikhin formula},
J. Phys. A: Math. Theor., {\bf 49}:45 (2016) 454005, 28 pp., \texttt{arXiv:1605.09189}.
%
\bibitem{HutLPRS16b} A.~A.~Hutsalyuk,  A.~Liashyk, S.~Z.~Pakuliak, E.~Ragoucy, N.~A.~Slavnov,
\textsl{Scalar products of Bethe vectors in models with $\mathfrak{gl}(2|1)$ symmetry 2. Determinant representation},
J. Phys. A: Math. Theor., {\bf 50}:3 (2017) 34004, 22 pp., \texttt{arXiv:1606.03573}.
%
\bibitem{PakRS15a} S. Pakuliak, E. Ragoucy, N. A. Slavnov, \textsl{Zero modes method and form factors in quantum integrable models},
Nucl. Phys. B, {\bf 893} (2015) 459--481, \texttt{arXiv:1412.6037}.
%
\bibitem{PakRS15c}  S. Pakuliak, E. Ragoucy, N. A. Slavnov, \textsl{$GL(3)$-Based Quantum Integrable Composite Models. II. Form Factors of Local Operators},
SIGMA {\bf 11} (2015) 064, \texttt{arXiv:1502.01966}.
%
\bibitem{HutLPRS16d} A.~A.~Hutsalyuk,  A.~Liashyk, S.~Z.~Pakuliak, E.~Ragoucy, N.~A.~Slavnov,
\textsl{Form factors of the monodromy matrix entries in $\mathfrak{gl}(2|1)$-invariant integrable models},
Nucl. Phys. B, {\bf 911} (2016), 902--927, \texttt{arXiv:1607.04978}.
%
\bibitem{FukS17} J.~Fuksa, N. A. Slavnov, \textsl{Form factors of local operators in supersymmetric quantum integrable models},
J. Stat. Mech. Theory Exp., (2017), 043106, \texttt{arXiv:1701.05866}.
%
\bibitem{PakRS14a} S. Z.  Pakuliak, E. Ragoucy, N. A. Slavnov, \textsl{Scalar products in models with the $GL(3)$ tri\-go\-no\-metric $R$-matrix: General case},
Theor. Math. Phys. {\bf 180}:1 (2014) 795--814, \texttt{arXiv:1401.4355}.
%
\bibitem{Sla15a} N. A. Slavnov, \textsl{Scalar products in $GL(3)$-based models with trigonometric $R$-matrix. Determinant representation}, J. Stat. Mech.
 (2015) P03019, \texttt{arXiv:1501.06253}.
%
\bibitem{VT}
V. Tarasov, A. Varchenko, {\sl Jackson inte\-gral re\-pre\-sen\-ta\-ti\-ons of so\-lu\-ti\-ons
of the quan\-tized Knizh\-nik--Za\-mo\-lod\-chi\-kov equation},  Algebra and Analysis, {\bf 6}:2 (1994) 90--137;
St. Petersburg Math. J. {\bf 6}:2 (1995) 275--313 (Engl. transl.), \texttt{arXiv:hep-th/9311040}.
%
\bibitem{TarV96} V. Tarasov, A. Varchenko, \textsl{Asymptotic Solutions to the Quantized Knizhnik-Za\-mo\-lod\-chi\-kov Equation and Bethe Vectors},
Amer. Math. Society Transl.,  Ser. 2 {\bf 174} (1996) 235--273, \texttt{arXiv:hep-th/9406060}.
%
\bibitem{MukV05} E. Mukhin,  A. Varchenko, \textsl{Norm of a Bethe vector and the Hessian of the master function},
Compositio Math. {\bf141} (2005) 1012--1028, \texttt{arXiv:math/0402349}.
%
\bibitem{WheelF13}
O. Foda, M. Wheeler,
\textsl{Colour-independent partition functions in coloured vertex models},
Nucl. Phys. \textbf{B871} (2013) 330 -- 361, \texttt{arXiv:1301.5158}.
%
\bibitem{EscGSV11} J. Escobedo, N. Gromov, A. Sever, P. Vieira, \textsl{Tailoring Three-Point Functions and Integrability},
JHEP 1109 (2011) 028, \texttt{arXiv:1012.2475}
%
\bibitem{Grom16} N. Gromov, F. Levkovich-Maslyuk, G. Sizov,
\textsl{New Construction of Eigenstates and Separation of Variables for SU(N) Quantum Spin Chains},
\texttt{arXiv:1610.08032}.
%
\bibitem{HutLPRS17a} A.~A.~Hutsalyuk,  A.~Liashyk, S.~Z.~Pakuliak, E.~Ragoucy, N.~A.~Slavnov,
\textsl{Current presentation for the double super-Yangian $DY(\mathfrak{gl}(m|n))$ and Bethe vectors},
Russ. Math. Surv. {\bf 72}:1  (2017), 33--99, \texttt{arXiv:1611.09020}.
%
\bibitem{KhP-Kyoto} S. Khoroshkin, S. Pakuliak, {\sl A computation of an universal weight function for
the quantum affine algebra $U_q(\mathfrak{gl}(N))$},  {J. of Mathematics of Kyoto University},
{\bf 48} n.2 (2008) 277--321, \texttt{arXiv:0711.2819}.
%
\bibitem{DiFr93}
J. Ding, I. B. Frenkel, {\sl Isomorphism of two realizations of quantum affine algebra
$U_q(\mathfrak{gl}(N))$}, Commun.~Math.~Phys. {\bf 156} (1993), 277--300.
%
\bibitem{Ize87} A. G. Izergin, \textsl{Partition function of the six-vertex model in a finite volume},
Dokl. Akad. Nauk SSSR {\bf 297} (1987) 331--333;
Sov. Phys. Dokl. {\bf 32} (1987) 878--879 (Engl. transl.).
%
\bibitem{EssK94} F. H. L. Essler, V. E. Korepin, \textsl{Spectrum of Low-Lying Excitations in a Supersymmetric
Extended Hubbard Model}, Int. J. Mod. Phys. B{\bf 8} (1994) 3243--3279, \texttt{arXiv:cond-mat/9307019}
%
\bibitem{For89} D. F\"orster, \textsl{Staggered spin and statistics in the supersymmetric t-J model}, Phys.
Rev. Lett. {\bf 63} (1989) 2140--2143.
%
\bibitem{EssK92} F. H. L. Essler and V. E. Korepin, \textsl{Higher conservation laws and algebraic Bethe
Ansatze for the supersymmetric t-J model}, Phys. Rev. B {\bf 46} (1992) 9147--9162.
%
\bibitem{FoeK93} A. Foerster and M. Karowski, \textsl{Algebraic properties of the Bethe ansatz for an
$spl(2,1)$-supersymmetric t-J model}, Nucl. Phys. B {\bf 396} (1993) 611--638.
%
\bibitem{Sch87} P. Schlottmann, \textsl{Integrable narrow-band model with possible relevance to heavy
Fermion systems}, Phys. Rev. B {\bf 36} (1987) 5177--5185.
%
\bibitem{BachFoer}
M. T. Batchelor, A. Foerster, \textsl{Yang-Baxter integrable models in experiments: from condensed matter to ultracold atoms},
J. Phys. A: Math. Theor. \textbf{49} (2016) 173001, \texttt{arXiv:1510.05810}.
%
\bibitem{KulS80} P. P. Kulish and E. K. Sklyanin, \textsl{On the solution of the Yang--Baxter equation},
Zap. Nauchn. Semin. LOMI {\bf 95} (1980) 129--160;  J. Sov. Math. {\bf 19} (1982) 1596--1620 (Engl. transl.).
%
\bibitem{PakRS17} S. Pakuliak, E.  Ragoucy, and N. A. Slavnov,
{\sl Bethe vectors for models based on the super-Yangian $Y(\mathfrak{gl}(m|n))$},
J. Integrable Systems {\bf 2} (2017)  1--31, \texttt{arXiv:1604.02311}.
%
\bibitem{Fuk17} J.~Fuksa, \textsl{Bethe vectors for composite generalised models
with $\mathfrak{gl}(2|1)$ and $\mathfrak{gl}(1|2)$ supersymmetry}, SIGMA {\bf 13} (2017), 015, 17pp., \texttt{arXiv:1611.00943}.
%
\bibitem{PakRS15d}  S. Pakuliak, E. Ragoucy, N. A. Slavnov, \textsl{$GL(3)$-Based Quantum Integrable Composite Models. I. Bethe vectors},
SIGMA {\bf 11} (2015) 063, \texttt{arXiv:1501.07566}.
%
\bibitem{alexbook} A. Molev, \textsl{Yangians and classical Lie algebras}, Math. Surveys and Monographs \textbf{143}, ed. Am. Math. Soc. (2007)
%
\bibitem{Sla16} N. A. Slavnov, \textsl{Multiple commutation relations in the models with $gl(2|1)$ symmetry},
Theor. Math. Phys., {\bf189}:2 (2016), 1624--1644, \texttt{arXiv:1604.05343}.
%
\bibitem{Frank} F. G\"oehmann, A. Seel,
\textsl{Algebraic Bethe ansatz for the $gl(1|2)$ generalized model II: the three gradings},
J. Phys. A \textbf{37} (2004) 2843, \texttt{cond-mat/0309135}.
%
\bibitem{AACRFE} D.~Arnaudon, J.~Avan, N.~Cramp\'e, A.~Doikou, L.~Frappat, and {\'E}.~Ragoucy,
\textsl{General boundary conditions for the $sl(N)$ and super $sl(M|N)$
open spin chains}, J. Stat. Mech. \textbf{0408} (2004) P005, \texttt{math-ph/0406021}.
%
\bibitem{Satta} E. Ragoucy and G. Satta,
\textsl{Analytical {B}ethe ansatz for closed and open $gl({M}|{N})$
super-spin chains in arbitrary representations and for any {D}ynkin
diagrams}, JHEP \textbf{0709} (2007) 001, \texttt{arXiv:0706.3327}.
%
\end{thebibliography}
\end{document}